\documentclass[10pt,letterpaper]{article}
\usepackage[top=0.85in,left=2.75in,footskip=0.75in]{geometry}

\usepackage{amsmath,amssymb}

\usepackage{changepage}

\usepackage[utf8x]{inputenc}

\usepackage{textcomp,marvosym}

\usepackage{cite}

\usepackage{nameref,hyperref}

\usepackage[right]{lineno}

\usepackage{microtype}
\DisableLigatures[f]{encoding = *, family = * }

\usepackage[table]{xcolor}

\usepackage{array}

\newcolumntype{+}{!{\vrule width 2pt}}

\newlength\savedwidth

\raggedright
\usepackage[aboveskip=1pt,labelfont=bf,labelsep=period,justification=raggedright,singlelinecheck=off]{caption}

\makeatletter
\renewcommand{\@biblabel}[1]{\quad#1.}
\makeatother

\usepackage{lastpage,fancyhdr,graphicx}
\usepackage{epstopdf}
\fancyheadoffset[L]{2.25in}
\fancyfootoffset[L]{2.25in}
\usepackage[graphicx]{realboxes}

\begin{document}
\vspace*{0.2in}

\begin{flushleft}
{\Large
\textbf\newline{Circuits with broken fibration symmetries perform core logic computations in
  biological networks} 
}
\newline
\\
Ian Leifer\textsuperscript{1},
Flaviano Morone\textsuperscript{1},
Saulo D. S. Reis\textsuperscript{2},
Jos\'e~S.~Andrade, Jr.\textsuperscript{2},
Mariano~Sigman\textsuperscript{3},
Hern\'an A. Makse\textsuperscript{1*}
\\
\bigskip
\textbf{1} Levich Institute and Physics Department, City College
  of New York, New York, New York, USA
\\
\textbf{2} Departamento de F\'isica, Universidade Federal do Cear\'a,
Fortaleza, Cear\'a, Brazil
\\
\textbf{3} Laboratorio de Neurociencia, Universidad Torcuato Di
Tella, Buenos Aires, Argentina
\\
\bigskip

%
%





* hmakse@ccny.cuny.edu

\end{flushleft}
\section*{Abstract}

We show that logic computational circuits in gene regulatory networks
arise from a fibration symmetry breaking in the network
structure. From this idea we implement a constructive procedure that
reveals a hierarchy of genetic circuits, ubiquitous across species,
that are surprising analogues to the emblematic circuits of
solid-state electronics: starting from the transistor and progressing
to ring oscillators, current-mirror circuits to toggle switches and
flip-flops. These canonical variants serve fundamental operations of
synchronization and clocks (in their symmetric states) and memory
storage (in their broken symmetry states). These conclusions introduce
a theoretically principled strategy to search for computational
building blocks in biological networks, and present a systematic route
to design synthetic biological circuits.

    \section*{Author summary}

    We show that the core functional logic of genetic circuits arises
    from a fundamental symmetry breaking of the interactions of the
    biological network. The idea can be put into a hierarchy of
    symmetric genetic circuits that reveals their logical
    functions. We do so through a constructive procedure that
    naturally reveals a series of building blocks, widely present
    across species.  This hierarchy maps to a progression of
    fundamental units of electronics, starting with the transistor,
    progressing to ring oscillators and current-mirror circuits and
    then to synchronized clocks, switches and finally to memory
    devices such as latches and flip-flops.

\linenumbers

\section*{Introduction}

In all biological networks \cite{leibler} some simple `network motifs'
appear more often than they would by pure chance
\cite{alon-motif,alon-ecoli,alon}.  This regularity has been
interpreted as evidence that these motifs are basic building blocks of
biological machineries, a proposal that has had a major impact on
systems biology \cite{alon,klipp}.  However, mere statistical
abundance by itself does not imply that these circuits are core bricks
of biological systems. In fact, whether these network motifs may have
a functional role in biological computation remains controversial
\cite{stumpf,wagner,macia,fink}.

Functional building blocks should offer computational repertoires
drawing parallels between biological networks and electronic circuits
\cite{tyson}. Indeed, the idea of using electronic circuitry and
devices to mimic aspects of gene regulatory networks has been in
circulation almost since the inception of regulatory genetics itself
\cite{monod}. This idea has been a driving force in synthetic biology
\cite{teo}; with several demonstrations showing that engineered
biological circuits can perform computations \cite{dalchau}, such as
toggle switches
\cite{atkinson2003,gardner2000,kramer_fussenegger2005,kramer2004},
logic \cite{guet2002}, memory storage
\cite{Ajo_Franklin2007,gardner2000,Ham2006}, pulse generators and
oscillators
\cite{Elowitz_Leibler2000,atkinson2003,stricker2008,Tigges2009}.

In a previous work \cite{fibration}, we have shown that the building
blocks of gene regulatory networks can be identified by the fibration
symmetries of these networks. In the present paper, we first
demonstrate the functionality of these symmetric building blocks in
terms of synchronized biological clocks.  We then show that the
breaking of the fibration symmetries of the network identifies
additional building blocks with core logic computational functions.
We do so through a constructive procedure based on symmetry breaking
that naturally reveals a hierarchy of genetic building blocks widely
present across biological networks and species.  This hierarchy maps
to a progression of fundamental units of electronics, starting with
the transistor, and progressing to ring oscillators and current-mirror
circuits and then to memory devices such as toggle switches and
flip-flops. We show that, while symmetric circuits work as
synchronized oscillators, the breaking of these symmetries plays a
fundamental role by switching the functionality of circuits from
synchronized clocks to memory units.


Thus, our constructive theoretical framework identifies 1) building
blocks of genetic networks in a unified way from fundamental symmetry
and broken symmetry principles, which 2) assure that they perform core
logic computations, 3) suggests a natural mapping onto the
foundational circuits of solid state electronics \cite{teo,dalchau},
and 4) endows the mathematical notion of symmetry fibration
\cite{fibration} with biological significance.

\section*{Results}

\subsection*{Feed-forward loops do not synchronize}
\label{main:ffl}

\begin{figure}[!h]
	\centering
	\includegraphics[width=0.95\linewidth]{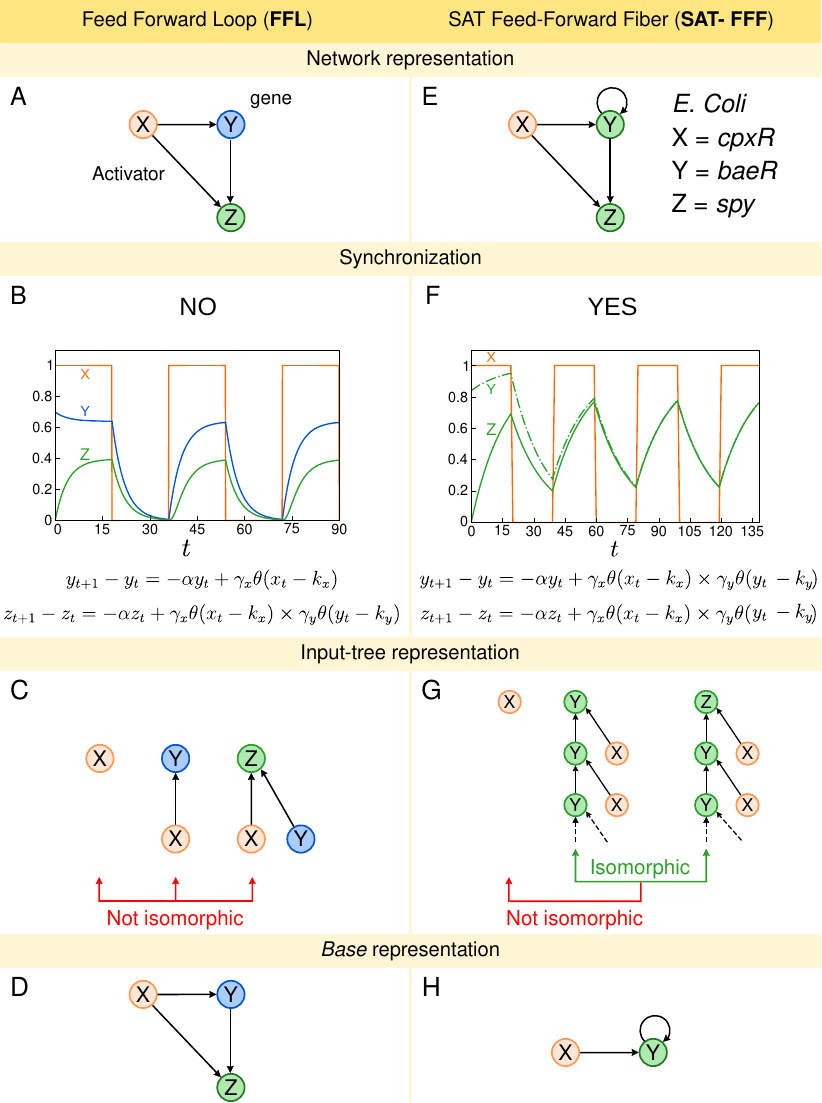}
\caption{\small {\bf Feed-Forward Loop (FFL) and Feed-Forward Fiber
  (FFF).}  A: FFL network representation. B: Numerical
solution of FFL dynamics.  The expression levels of genes Y and Z do
not synchronize. The oscillation pattern presented is due to the
square-wave behavior of gene X expression levels. We use $\alpha =
2.0$, $\gamma_x = 0.12$, $\gamma_y= 0.7$, $k_x = 0.5$, $k_y=0.1$, $y_0
= 0.7$ and $z_0 = 0.0$. C: Input tree representation of FFL. The
input trees of genes X, Y, and Z are not isomorphic, as a consequence,
their expression levels do not synchronize. D: Base
representation of FFL. The {\it base} is the same as the original
circuits since there are no symmetries. E: SAT-FFF network
representation. As an example, we consider genes {\it cpxR} = X, {\it
  baeR} = Y, and {\it spy} = Z from the gene regulatory network of
{\it E. coli}. The addition of the autoregulation leads to a symmetry
between the expression levels of genes Y and Z. F: The numerical
solution of the SAT-FFF dynamics shows the synchronization of the
expression levels of genes Y and Z. We use $\alpha = 0.06$, $\gamma_x
= 0.775$, $\gamma_y=0.775$, $k_x = 0.5$, $k_y = 0.1$, $y_0 = 0.85$ and
$z_0 = 0.0$.  Again, the oscillation is due to the wave-like pattern
of X. G: As a result, genes Y and Z have isomorphic input
trees. However, the input tree of the external regulator {\it cpxR} is
not isomorphic, despite the fact that it directly regulates the
\textit{fiber}.  H: Since Y and Z synchronize, gene Z can be
collapsed into Y, resulting in a simpler {\it base} representation.}
\label{fig1}
\end{figure}

We start our analysis by considering the dynamics of the most abundant
network motif in transcriptional regulatory networks, the so-called
feed-forward loop (FFL) introduced by Alon and coworkers
\cite{alon-ecoli,alon,mangan}.  A cFFL motif consists of three genes
(X, Y and Z; c refers to coherent where all regulators are activators)
where the transcription factor expressed by gene X positively
regulates the transcription of Y and Z, and, in turn Y regulates Z
(Fig. \ref{fig1}A).  Fig.~\ref{fig1}A shows an example of cFFL motif
in {\it E. coli} with X={\it cpxR}, Y={\it baeR}, and Z={\it spy}.
Numerical and analytic solutions for the expression levels of the
genes in the cFFL (Fig.~\ref{fig1}B)
demonstrate that the FFL does not reach synchronization
(unless for very specific setting of parameters) nor
oscillations in expression levels.  This is consistent with previous
research which has interpreted the functionality of the FFL as signal
delay~\cite{alon-ecoli,mangan,alon-delay}.

We illustrate this result by presenting an analytical
  solution of the FFL \cite{alon-motif,alon-ecoli,alon}.
We use a discrete time, continuous state variable model with a logic
Boolean interaction function in the spirit of the Glass and Kauffman
model of biochemical networks~\cite{kauffman}. The dynamics of the
expression levels $y_t$ and $z_t$ of genes Y and Z, respectively, as a
function of time $t$ in the cFFL is given by the following difference
equations \cite{alon}:
\begin{equation}
\begin{aligned}
y_{t+1} &= (1-\alpha) y_t + \gamma_x \theta(x_t-k_x),\\
z_{t+1} &= (1-\alpha) z_t + \gamma_x \theta(x_t-k_x) \times \gamma_y \theta(y_t-k_y),
\end{aligned}
\label{eq:fflbool}
\end{equation}
where $x_t$ is the expression level of gene X, $\alpha$ is the
degradation rate of the gene, $\gamma_x$ and $\gamma_y$ are the
strength of the interaction representing the maximum expression rate
of genes X and Y, respectively, and the thresholds $k_x$ and $k_y$ are
the dissociation constant between the transcription factor and biding
site. The expression level is measured in terms of abundance of gene
product, e.g., mRNA concentration.  The Heaviside step functions
$\theta(x_t-k_x)$ and $\theta(y_t-k_y)$ represent the activator
regulation from gene X and Y, respectively. They represent the Boolean
logic approximation of Hill input functions in the limit of strong
cooperativity \cite{kauffman,alon}.  We consider an AND gate for the
combined interaction of transcription factors of genes X and Y onto
the binding sites of gene Z \cite{alon}.  Analogous results shown in
S1 File Section~I can be obtained with an OR gate and with ODE
continuum models.

In S1 File Sec.~I~A, we show that the expression levels of the genes
Y and Z do not synchronize, in other words, $y_t \neq z_t$, $\forall t$.
For example, Fig.~\ref{fig1}B shows a particular set of parameters which results
in a non-synchronized state. Such state is obtained under initial condition  
$y_0>k_y$
and $\alpha<\gamma_x$. Specifically, we use the parameters: $\alpha =
0.2$, $\gamma_x = 0.12$, $\gamma_y = 0.7$, $k_x = 0.5$ and $k_y =
0.1$.  For this combination, $y_t$ and $z_t$ do not synchronize since
$y_t$ saturates at $y_t \to \gamma_x/\alpha = 0.6$ when $t \to
\infty$, and $z_t$ saturates at $z_t \to \gamma_x\gamma_y/\alpha =
0.42$, for $t \to \infty$. In this figure, we set $x_t$ equal to a
square wave and then monitor the expression levels of $y_t$ and
$x_t$. When $x<k_x$, both $y_t$ and $z_t$ decay exponentially to
zero. On the other hand, when $x>k_x$, both variables evolve to
saturate again at $y_t = \gamma_x/\alpha$ and $z_t =
\gamma_x\gamma_y/\alpha$, in agreement with the analytical solution.


\subsection*{Feed-forward fibers synchronize via a symmetry fibration}

In the FFL, gene Z receives input from X and Y, while gene Y, instead,
only from X, and therefore the inputs are not symmetric
(Fig.~\ref{fig1}C).  But as it turns out, a search of motifs in
biological networks \cite{fibration} shows that the FFL circuit in
Fig. \ref{fig1}A regularly appears in conjunction with an
autoregulation (AR) loop \cite{alon-autoregulation} at Y={\it baeR}
(Fig. \ref{fig1}E).  This minimal inclusion in the FFL symmetrizes the
circuit and we show, next, that this symmetry results in a circuit
with a first and minimal form of function: synchronization in gene
expression.  This can be formalized by analyzing invariances in the
input tree \cite{fibration}, which represents all the paths that
converge to a given gene (Fig.~\ref{fig1}C and G).  The mathematical
principle of symmetry fibration in dynamical systems introduced in
~\cite{fibration,golubitsky,boldi} predicts this, since symmetries Y
$\leftrightarrow$ Z that leave invariant the tree of inputs (but not
necessarily the outputs as imposed by automorphisms) are necessary and
sufficient to achieve synchronization. When two genes have isomorphic
input trees, their expression levels are synchronized, see
~\cite{fibration} for details.



For instance, 
by means of its autoregulation, {\it baeR} forms a symmetry fibration with gene
{\it spy}, that can be formalized and characterized by an isomorphism
between the input trees of these genes (Fig. \ref{fig1}G). When two
input trees, like those of {\it baeR} and {\it spy}, are isomorphic,
the expression levels of these two genes are synchronized. 
These genes are said to belong to the same
\textit{`fiber'}~\cite{boldi,fibration}.  Genes in the same
\textit{fiber} are redundant, and can be collapsed into the
`\textit{base}' (see Fig.~\ref{fig1}H) by a symmetry
fibration~\cite{fibration} (the FFL, instead, cannot be reduced
since it has no symmetry, see Fig.~\ref{fig1}D). Numerical simulations
and analytical solutions (Fig.~\ref{fig1}F, S1 File
Section~II A and Section~II~B) confirm that the
addition of the AR loop to the FFL -- leading to a circuit that we
call the Feed-Forward Fiber (FFF) -- changes its functionality
qualitatively, leading to synchronization of genes Y and Z into
coherent co-expression.  This prediction is confirmed with
experimental co-expression profiles in Ref.~\cite{fibration}.




The FFF with activator regulations in Fig.~\ref{fig1}E has a simple dynamics
converging to a synchronized fixed point: all interactions are satisfied meaning
that the Heaviside step functions evaluate to 1 and we call this circuit SAT-FFF. 
Instead, when the autoregulation is a repressor, the
loop behaves as a logical NOT gate.  When expression is high
it inhibits itself shifting to low state. Instead, if it is low it promotes
itself to shift to a high state. Hence, the activity of this gene oscillates
indefinitely~\cite{atkinson2003,kauffman}. This is the simplest expression of 
{\it frustration} ~\cite{anderson,kauffman},
a core concept in physics which refers to a system which is always in
tension and thus never reaches a stable fixed configuration.

\subsection*{A biological transistor as a core building block}







\begin{figure}[!h]
	\centering
	\includegraphics[width=0.95\linewidth]{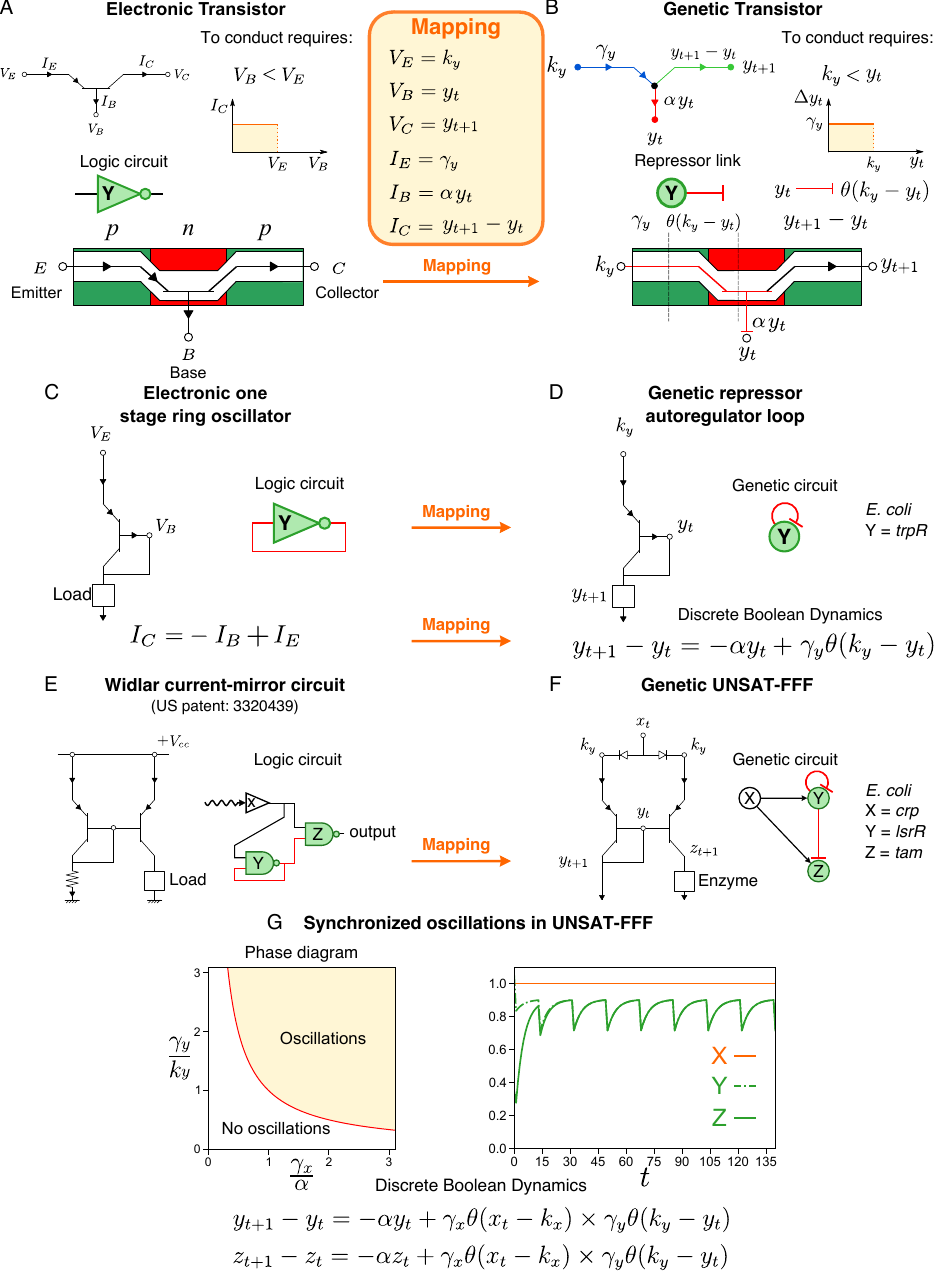}
\caption{\small {\bf Mapping between electronic and biological
  transistor.}  A: A pnp transistor allows current flow if the
voltage applied to its base is lower than the voltage at its emitter
($V_B<V_E$). Since it has a high (low) output for a lower (high)
input, it is logically represented by a NOT gate.  The yellow box
shows the mapping between the pnp transistor and the biological
repressor. B: A repressor regulation link plays the role of the
pnp transistor since the rate of expression of a gene is repressed
by gene Y if $k_y<y_t$. C: By connecting the base of the
transistor to its collector, one forms a one stage ring oscillator. D:
This connection is translated to the biological analogue as a repressor
autoregulation at gene Y.  In this way, the rate expression of gene Y
is able to oscillate, depending on the adjustment of parameters
$\alpha$, $k_y$ and $\gamma_y$ (see S1 File Section~III).
E: Widlar current-mirror circuit and F: its biological
analogue (UNSAT-FFF). By mirroring the ring oscillator, the Widlar
mirror circuit allows synchronization and oscillations (see S1 File
Section~III). G: Phase diagram of oscillations
of the UNSAT-FFF. An oscillatory phase is defined by the
condition $\gamma_y/k_y > \left(\gamma_x/\alpha\right)^{-1}$ (see S1 File
Section~III). For instance, on the right side, we plot
the solution of the discrete dynamics for a set of parameters
satisfying such condition. Specifically, $\alpha=0.205$, $\gamma_x =
0.454$, $\gamma_y=0.454$, $k_x=0.5$, and $k_y=1.0$.}
\label{fig2}
\end{figure}

To understand the computation rationale of symmetric and frustrated
circuits made of repressors, we map them to electronic analogues.  We
begin the analogy with the simplest circuit of a single gene with a
feedback loop with repression (AR loop, Fig.~\ref{fig2}D). The
dynamics for the expression level $y_t$ is described by the discrete
time model with Boolean interaction \cite{kauffman}:

\begin{equation} \Delta y_t = y_{t+1} - y_t = -\alpha y_t
  + \gamma_y\theta(k_y-y_t).
\label{eq:analogue}
\end{equation}
Here, $\alpha$ is the degradation rate of the Y gene product,
$\gamma_y$ is the maximum expression rate of gene Y, and $k_y$ is the
dissociation constant.

The
Heaviside step function $\theta(k_y-y_t)$ reflects the repressor
autoregulation in the Boolean logic approximation.
We will show that this genetic repressor interaction, shown as the stub in Fig.~\ref{fig2}B,
is the genetic analogue of a solid-state transistor shown in Fig.~\ref{fig2}A.


A transistor is typically made up
of three semiconductors, a base sandwiched between an emitter and a
collector (Fig.~\ref{fig2}A).  The current flows between the emitter
and collector only if voltage applied to the base is lower than at the
emitter ($V_B<V_E$) and thus the transistor acts as a switch and
inverter. In the genetic circuit, the expression $y_t$ drives
the rate of expression of gene Y, like the voltage drives current
around an electric circuit.  Simply comparing Eq. (\ref{eq:analogue})
to the pnp transistor in Fig. \ref{fig2}A leads to the analogy in
which the expression $y_t$ is an analogue for the base potential
$V_B$ of a transistor, $k_y$ an analogue for $V_E$, $\gamma_y$ an
analogue for the emitter current $I_E$, $\alpha y_t$ an analogue for
$I_B$, and $\Delta y_t$ an analogue for $I_C$. Then,
Eq.~\eqref{eq:analogue} provides the genetic equivalent of the
equation for a transistor's collector current $I_C = I_E - I_B$
(Fig. \ref{fig2}C, D).

The repressor AR genetic circuit of Fig.~\ref{fig2}D becomes a one-stage ring oscillator
(Fig.~\ref{fig2}C) where the collector of the transistor connects to
the base forming the minimal signal feedback loop. As shown in 
Fig.~\ref{fig2}D, an example of the AR is the gene Y~=~{\it trpR} from
the {\it E. coli} transcriptional network. The repressor AR loop can be
extended to the FFF by symmetrizing
it, adding gene Z, such that it synchronizes with Y to express an enzyme
that catalyzes a biochemical reaction (Fig.~\ref{fig2}F). The circuit is
completed with the external regulator X which keeps the symmetry between Y
and Z. The resulting circuit is called UNSAT-FFF (since it is frustrated).


The UNSAT-FFF maps to the so-called Widlar current-mirror
electronic circuit shown in Fig. \ref{fig2}E, a popular building block of
integrated circuits used since the foundations of the semiconductor
industry (1967 US patent \cite{widlar,horowitz}). It serves two key
functions as we show below: mirror synchronization of $y_t=z_t$ and oscillatory
activity (Fig. \ref{fig2}G and S1 File Section~III).

\subsection*{UNSAT-FFF solution oscillates and synchronize}
\label{main:unsat}

Next, we show analytically and numerically that the UNSAT-FFF circuit
has an oscillatory solution plus synchronization of genes Y and Z. S1
File Section~III~A shows that the prediction of oscillatory behavior
can be obtained also from a model of gene expression in the continuum
time-domain using a first-order ODE model with time delay due to the
process of transcription and translation. Below, we focus on the
discrete dynamics.

The UNSAT-FFF circuit is obtained by the addition of a repressor AR
loop to the FFL: in Fig.~\ref{fig2}F, gene Y acts as a repressor
regulator on the gene Z and on itself.  There are many variants of
this circuit depending on the combinations of activator and repressor
regulators. Here, we use X gene as an activator and Y gene as a
repressor. Different variants yields analogous results and will be
discussed elsewhere.  The important ingredient is the existence of
frustration in the interactions. For instance if gene X is high, then
it will make genes Y and Z high too. However, this configuration does
not satisfy the repressor autoregulation bond neither the repression
from Y~$\to$~Z.  Thus, two bonds are unsatisfied and this circuit
unsatisfied: UNSAT-FFF. The discrete-time dynamics of the expression
levels of genes $y_t$ and $z_t$ are given by:
\begin{equation}
\begin{aligned}
y_{t+1} &= (1-\alpha) y_t + \gamma_x \theta(x_t-k_x)\times \gamma_y \theta(k_y-y_t)\ ,\\
z_{t+1} &= (1-\alpha) z_t + \gamma_x \theta(x_t-k_x) \times \gamma_y \theta(k_y-y_t)\ ,
\end{aligned}
\label{eq:fffnegative}
\end{equation}
where $\gamma_x$ and $\gamma_y$ are the strength of the interaction
(maximum expression rate) of genes $X$ and $Y$, respectively, and
$k_x$ and $k_y$ are the dissociation constant, respectively.
Similarly to the SAT-FFF case, synchronization between $y$ and $z$
occurs for the UNSAT-FFF (see S1 File Section~II).
However, the impact of the repressor feedback loop on the dynamical
behavior of this circuit is more profound, since it leads to
oscillations. Thus, while both, SAT-FFF and UNSAT-FFF, lead to
synchronization of Y and Z, the former synchronizes into a fixed point
and the later into an oscillatory limit cycle.

We set $\lambda=\gamma_x\gamma_y/k_y\alpha$, and $\beta = 1-\alpha$,
so that we rewrite Eq.~(\ref{eq:fffnegative}) for the rescaled
variables $\psi_t = y_t/k_y$ and $\zeta_t = z_t/k_y$ as
\begin{equation}
\begin{aligned}
\psi_{t+1} &= \beta\psi_t + \alpha\lambda\theta(x_t-k_x)\theta(k_y-\psi_t),\\
\zeta_{t+1} &= \beta\zeta_t + \alpha\lambda\theta(x_t-k_x)\theta(k_y-\psi_t).
\end{aligned}
\label{eq:fffnegativepsi}
\end{equation}
Now, we set $x_t = x$ constant in time for simplicity. For $x<k_x$,
the solutions exponentially decay as $\psi_t = \psi_0e^{-t/\tau}$ and
$\zeta_t = \zeta_0e^{-t/\tau}$, where $\psi_0$ is the initial
condition. For $x>k_x$, Eq.~(\ref{eq:fffnegativepsi}) defines an
iterative map which satisfies the following recursive equation:
\begin{equation}
f^t(\psi) = f^{t-1}(\beta\psi)\theta(\psi-1) + f^{t-1}(\beta\psi+\alpha\lambda)\theta(1-\psi).
\label{eq:unsatfffmap}
\end{equation}
This iterative map results in different solutions depending on the
value of $\lambda$.

We consider first the case where the initial condition is $\psi_0>1$.
Thus, the solution of Eq.~\eqref{eq:fffnegativepsi} is $\psi_t =
\psi_0e^{-t/\tau}$, where $\tau^{-1}=-\log(1-\alpha)$. This solution
is correct as long as $\psi_t>1$, but ceases to be valid at a certain
time $t^*$ such that $\psi_t<1$, which is given by $t^*=\lceil
\tau\log \psi_0\rceil$.

Next, we consider the case $\psi_0<1$. In this case the solution is
given by $\psi_t = \psi_0e^{-t/\tau} + \lambda(1-e^{-t/\tau})$, which
is always valid for $\lambda<1$. Thus, when $\lambda<1$ the system
does not oscillate but it converges monotonically to a fixed point
$\psi_\infty = \lambda$.  However when $\lambda>1$, this solution
ceases to be valid at the time $t^* = \lceil \tau\log \frac{\lambda
  -\psi_0}{\lambda -1}\rceil$ such that $\psi_t>1$. Therefore, the
solution $\psi_t$ oscillates in time for $\lambda>1$. For the case of
$\psi_0>1$, the explicit solution is given by the general analytical
expression which is plotted in Fig. \ref{fig2}G, right:
\begin{equation}
\begin{aligned}
\psi_t &= \psi_0e^{-t/\tau}\ \ \ &{\rm for}\ t\in\Big\{0,1, \hdots,t_1=\left\lceil \tau\log \psi_0\right\rceil\Big\},\\
\psi_t &= \psi_1e^{-(t-t_1)/\tau}+\lambda\left(1 - e^{-(t-t_1)/\tau}\right)\ \ \ 
&{\rm for}\ t\in\left\{t_1,\hdots,t_2=t_1+\left\lceil \tau\log \frac{\lambda-\psi_1}{\lambda-1}\right\rceil\right\},\\
\psi_t &= \psi_2e^{-(t-t_2)/\tau}\ \ \ &{\rm for}\ t\in\Big\{t_2,\hdots,t_3=t_2+\lceil \tau\log \psi_2\rceil\Big\}.\\
\end{aligned}
\label{smeq:evol_unsatfff}
\end{equation}
The general solution with initial condition $\psi(t_0)<1$ can 
be written in a similar way. 

Thus, the main condition for oscillations in the circuits is $\lambda>1$,
and if $\lambda<1$, there is no oscillatory behavior, and
the solution $\psi_t$ converges monotonically to $\lambda$.
Therefore, the oscillatory phase is separated from the non-oscillatory
phase by the condition:
\begin{equation}
\frac{\gamma_y}{k_y} = \left(\frac{\gamma_x}{\alpha}\right)^{-1},
\end{equation}
which is depicted in the phase diagram in Fig.~\ref{fig2}G, left.

Thus, the repressor autoregulation at gene Y converts the circuit into
a synchronized clock, a primary building block in any logic
computational device.
S1 File~Section~V and S2 File show all FFFs found across gene
regulatory networks of the studied species spanning {\it A. thaliana},
{\it M. tuberculosis}, {\it B. subtilis}, {\it E. coli}, {\it
  salmonella}, {\it yeast}, mouse and humans. We also show in
Table~\ref{symm_table} the count of circuits across species and their
associated Z-scores showing that these circuits are statistically
significant. The algorithm to find these {\it fibers} in biological
networks is explained in \cite{fibration} and S1 File Section~IV and
it is available at \url{https://github.com/makselab/FiberCodes}.

  \begin{table}[ht]
  \begin{adjustwidth}{-2.5in}{0in} 
  \caption{{\bf Symmetric circuits (\textit{fibers}) count \cite{fibration}.} }
   \resizebox{8in}{!}{
      \centering
      \begin{tabular}{|c | c c c | c c c | c c c | c c c | c c c|}
      \hline
        Species & Database & Nodes & Edges & & AR Fiber & & & FFF & & & Fibonacci Fiber & & & n = 2 Fiber & \\
        & & & & $N_{\rm real}$ & $N_{\rm rand} \pm SD$ & Z-score & $N_{\rm real}$ & $N_{\rm rand} \pm SD$ & Z-score & $N_{\rm real}$ & $N_{\rm rand} \pm SD$ & Z-score & $N_{\rm real}$ & $N_{\rm rand} \pm SD$ & Z-score \\ 
      \hline
        Arabidopsis Thaliana & ATRM & 790 & 1431 & 2 & $0.2 \pm 0.5$ & 4 & 2 & $0 \pm 0$ & Inf & 5 & $0.3 \pm 0.6$ & 8.1 & 0 & N/A & N/A \\
      \hline
        Micobacterium tuberculosis & Research article & 1624 & 3212 & 11 & $0.7 \pm 0.8$ & 13.2 & 6 & $0.2 \pm 0.4$ & 14.6 & 4 & $1.7 \pm 1.4$ & 1.7 & 0 & N/A & N/A \\
      \hline
        Bacillus subtilis & SubtiWiki & 1717 & 2609 & 35 & $0.3 \pm 0.5$ & 64.6 & 13 & $0.3 \pm 0.5$ & 23.4 & 1 & $1.3 \pm 1.2$ & -0.2 & 2 & $0 \pm 0$ & 63.2 \\
      \hline
        Escherichia coli & RegulonDB & 879 & 1835 & 14 & $0.2 \pm 0.5$ & 29.1 & 12 & $0.1 \pm 0.2$ & 49.4 & 2 & $0.5 \pm 0.8$ & 1.9 & 1 & $0 \pm 0$ & $>3$ \\
      \hline
        Salmonella SL1344 & SalmoNet & 1622 & 2852 & 21 & $0.7 \pm 0.8$ & 25 & 14 & $0.2 \pm 0.4$ & 32 & 2 & $1.4 \pm 1.3$ & 0.5 & 3 & $0 \pm 0$ & $>3$ \\
      \hline
        Yeast & & & & & 10 & & & 5 & & & 3 & & & 0 &  \\
        & YTRP\_regulatory & 3192 & 10947 & 10 & $0.3 \pm 0.6$ & 17.3 & 4 & $0.2 \pm 0.4$ & 8.5 & 2 & $1.8 \pm 1.3$ & 0.2 & 0 & N/A & N/A \\
        & YTRP\_binding & 5123 & 38085 & 2 & $0.1 \pm 0.3$ & 6.3 & 0 & N/A & N/A & 0 & N/A & N/A & 0 & N/A & N/A \\
      \hline
        Mouse & TRRUST & 2456 & 7057 & 1 & $0.1 \pm 0.4$ & 2.3 & 0 & N/A & N/A & 6 & $0.3 \pm 0.6$ & 9.3 & 0 & N/A & N/A \\
      \hline
        Human & & & & & 1 & & & 1 & & & 100 & & & 1 &  \\
        & TRRUST & 2718 & 8215 & 0 & N/A & N/A & 0 & N/A & N/A & 10 & $0.4 \pm 0.6$ & 16.3 & 0 & N/A & N/A \\
        & TRRUST\_2 & 2862 & 9396 & 0 & N/A & N/A & 0 & N/A & N/A & 11 & $0.4 \pm 0.7$ & 16 & 0 & N/A & N/A \\
        & KEGG & 5164 & 59680 & 1 & $0.06 \pm 0.25$ & 3.76 & 1 & $0 \pm 0$ & $>3$ & 79 & $0.6 \pm 0.7$ & 112 & 1 & $0 \pm 0$ & $>3$ \\
      \hline
    \end{tabular}}
    \begin{flushleft} We
      report the Z-scores showing that all found fibers are
      statistically significant. We use a random null model with the
      same degree sequence
      (and sign of interaction) as
      the original network to calculate the random count $N_{\rm
        rand}$ and compare with the real circuit count $N_{\rm real}$
      to get the Z-score.
\end{flushleft}
    \label{symm_table}
  \end{adjustwidth}
  \end{table}

\subsection*{A broad class of logic and dynamic repertoires in the class of fibers}

\begin{figure}[!h]
	\centering
	\includegraphics[width=0.95\linewidth]{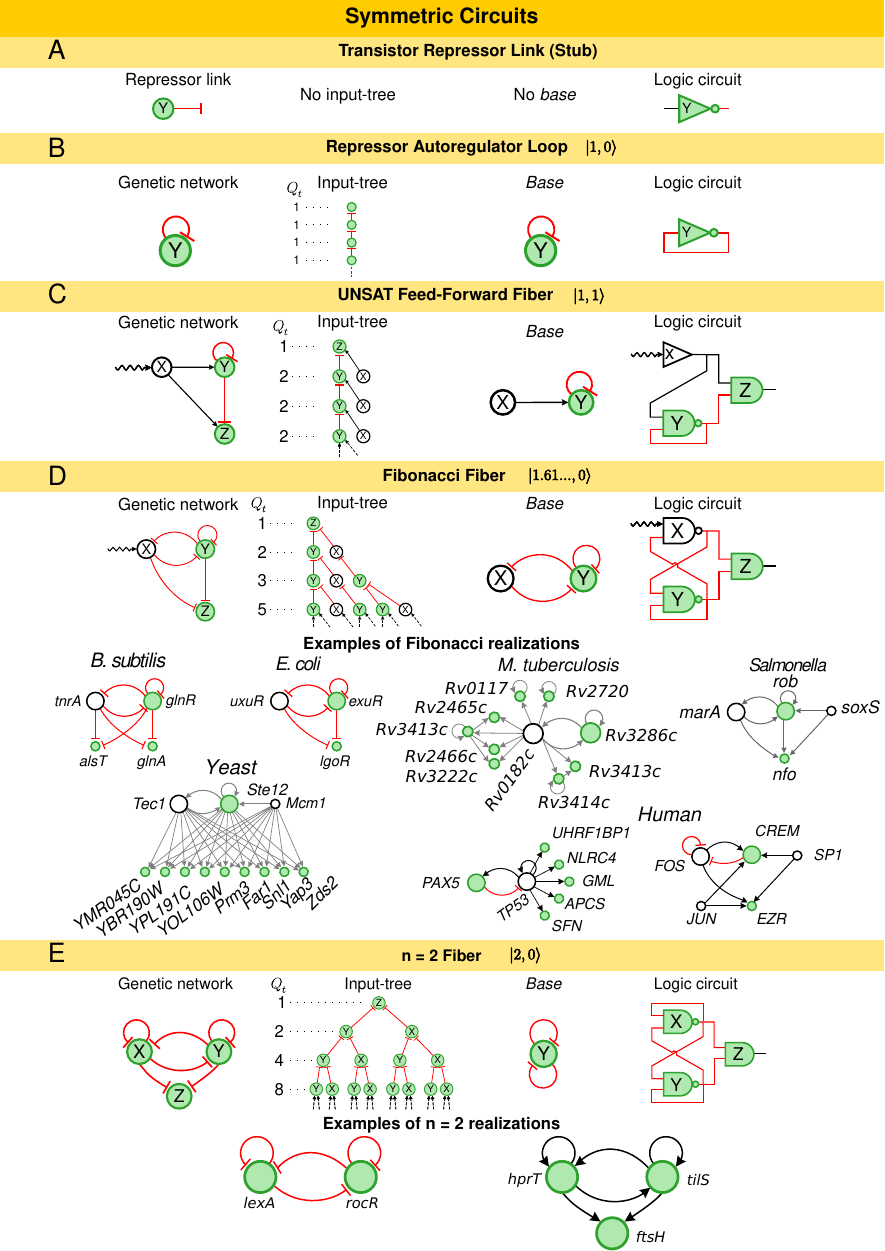}
\caption{\small {\bf Symmetric circuits \cite{fibration} function as clocks.}
  The addition of autoregulation loops and feedback loops results on a
  hierarchy of circuits of increasing complexity. For example, turning
  the A: repressor link into a B: repressor autoregulation results on
  an input tree that feeds its own expression levels with $Q_t=1$ and
  branching ratio $n=1$ and it is equivalent to its own {\it base}.
  C: The addition of an external regulator ($\ell=1$) creates the
  UNSAT-FFF where genes Y and Z synchronize and oscillate. This can be
  verified in its corresponding input tree and logic circuit.  D: The
  addition of a second feedback loop results in an input tree that
  follows the Fibonacci sequence $Q_t=1,2,3,5,8,...$. Here, gene X is
  not part of the {\it base}. The branching structure of the input
  tree implies that the Fibonacci Fiber can oscillate and synchronize,
  but is unable to store static information. Examples are shown from
  all studied species.
%
E: The second autoregulation at X results in a symmetric
input tree with $Q_t = 2Q_{t-1}$ and branching ratio $n=2$.  This
\textit{fiber} collapses into a {\it base} with two autoregulators.
Examples of $n=2$ Fiber are two \textit{fibers} from the regulatory
networks of {\it B. subtilis}.  In this figure, we present activator
links (black), repressor links (red), and interactions with unknown
functionality (grey).}
\label{fig3}
\end{figure}

The procedure to build more complex fibers can be systematically
extended through an algebra of circuits that adds external regulators
and loops to grow the {\it base} of symmetric circuits
(Fig. \ref{fig3} and S1 File Section~IV~A). In this space, the AR is
the core loop unit referred to as $|n=1, \ell=0\rangle$ in the
nomenclature of \cite{fibration} since it has $n=1$ loop and $\ell=0$
external regulator (Fig.~\ref{fig3}B), and the FFF is $|n=1,
\ell=1\rangle$ (Fig. \ref{fig3}C).

From this starting point, one can grow the number of regulator genes,
$| 1, \ell\rangle$ with $\ell > 1$. This does not affect the
complexity because all the relevant dynamics remain constrained to the
sole loop in the FFF circuit. {
Likewise, there are a
  number of circuits that can correspond to $|1, 1\rangle$. However,
  only certain modifications conserve the topological class identified
  by $|1, 1\rangle$. For instance, changing the sign of the edges is
  allowed as long as the edges of each gene are the same, but removing
  the edge X$\to$Z will break the symmetry of the fiber, so it is not
  allowed.  Adding a second node downstream of Y will conserve the
  topological class but only if it interacts with X.}  This situation
changes as soon as the \textit{fiber} feeds information back to the
external world.  This is the case of the circuits in Fig. \ref{fig3}D,
where the gene Y now regulates its own regulator gene X.  The
inclusion of this second feedback loop
results in the coexistence of two time-scales in the network.
{
This, in turn, increases the diversity of trajectories
  and delays in the network; a dynamic complexity that is measured
  by
the divergence of the input tree, 
  which is captured by the sequence $Q_t$ representing the number of
  source genes with paths of length $t - 1$ to the target gene, see
  Fig. \ref{fig3} and \cite{fibration}.  Simply put, the input tree is
  a rooted tree with a gene at the root ($Q_1=1$), and $Q_t$
  represents the number of genes in the $t-$th layer of the input
  tree.  Then, the divergence of the input tree is captured by its
  branching ratio measured by $n =
  \left(\frac{Q_{t+1}}{Q_t}\right)_{t\to\infty}$, see \cite{fibration}
  for details.}

A quantitative analysis of this measure yields exactly the golden
ratio~~~$\left(\frac{Q_{t+1}}{Q_t}\right)_{t\to\infty}=\varphi =
\left(1+\sqrt{5}\right)/2 = 1.6180...$ for the circuits in
Fig.~\ref{fig3}D, revealing that the input tree is a Fibonacci
sequence $Q_t= Q_{t-1} + Q_{t-2}$ updating the current state two steps
backwards, see \cite{fibration}.  We have called this class of
circuits, Fibonacci Fibers, in \cite{fibration}. For example, the
repressor interactions between genes X~=~{\it uxuR}, Y~=~{\it exuR},
and Z~=~{\it lgoR} from the {\it E. coli} network function exactly as
a Fibonacci Fiber (Fig.~\ref{fig3}D). Other typical examples of
Fibonacci Fibers in the transcriptional networks across species are
also shown in Fig.~\ref{fig3}D.  S1 File Section~V and S2 File show
all found {\it fibers}, and Table~\ref{symm_table} the counts for all
\textit{fibers} and their associated Z-scores showing that these
circuits are statistically significant. The important component of
these circuits is the delay in the feedback loop through the regulator
from Y~$\rightarrow$~X and back to Y captured by the $Q_{t-2}$ term in
the Fibonacci sequence.  This circuit has been
  synthetically implemented by Stricker {\it et al.}
  \cite{stricker2008} using a hybrid promoter that drives the
  transcription of genes {\it araC} and {\it lacI} forming a
  dual-feedback circuit. The functionality of this circuit has been
  demonstrated to be robust oscillations due to the negative feedback
  loop \cite{stricker2008}.  We will show next that a symmetry
  breaking in this Fibonacci circuit forms the {\it base} of the JK
  flip-flop, which is the universal storage device of computer
  memories ~\cite{horowitz}.


The complexity of the Fibonacci Fiber with feedback to the regulator is
$1.6180...$, which is lower than the number of loops in the circuit
(two). The intuition of what this reveals is that, in this circuit the
regulator X is still not part of the {\it base} (see Fig.~\ref{fig3}D), since
it does not receive input from itself and hence it is not within the
symmetry of the \textit{fiber}. This, in turn, indicates naturally that the next
element in the hierarchy of \textit{fibers} results from the inclusion of an AR
loop in X.  This creates a fully symmetric circuit (Fig.~\ref{fig3}E)
with a core $|2,\ell=0\rangle$ that feeds the reporter/enzyme Z. In this
case, the complexity of the circuit is exactly two (the number of
autoregulators within the \textit{fiber}).
Examples are shown in Fig.~\ref{fig3}E and Supplementary Materials.



\subsection*{Broken symmetry circuits as memory storage devices}

All symmetric circuits shown in Fig. \ref{fig3} \cite{fibration} work
as clocks with varying levels of sophistication and robustness given
by their time-scales or loops (see S1 File Sections~III~C and
IV~A). Additionally, the more complex Fibonacci Fibers store memory
dynamically by integrating sequences of its two immediate past states,
according to the input tree,
which computes temporal convolutions.  However, this is only a
short-term memory, stored dynamically and continually erased.  This
raises the necessity of understanding how these canonical biological
circuits can perform controlled memory set and reset, a fundamental
component of all computing devices~\cite{horowitz}.

We show next that static memory storage requires breaking the
fibration symmetry of each circuit creating structures analogous to
`flip-flops'~\cite{horowitz} in electronics that use a bistable toggle
switch~\cite{atkinson2003,gardner2000} to store a bit of binary
information into computer memory.
As we show below, in genetic networks, symmetry breaking endows the
circuits with the ability to remember.

\begin{figure}[!h]
	\centering
	\includegraphics[width=0.95\linewidth]{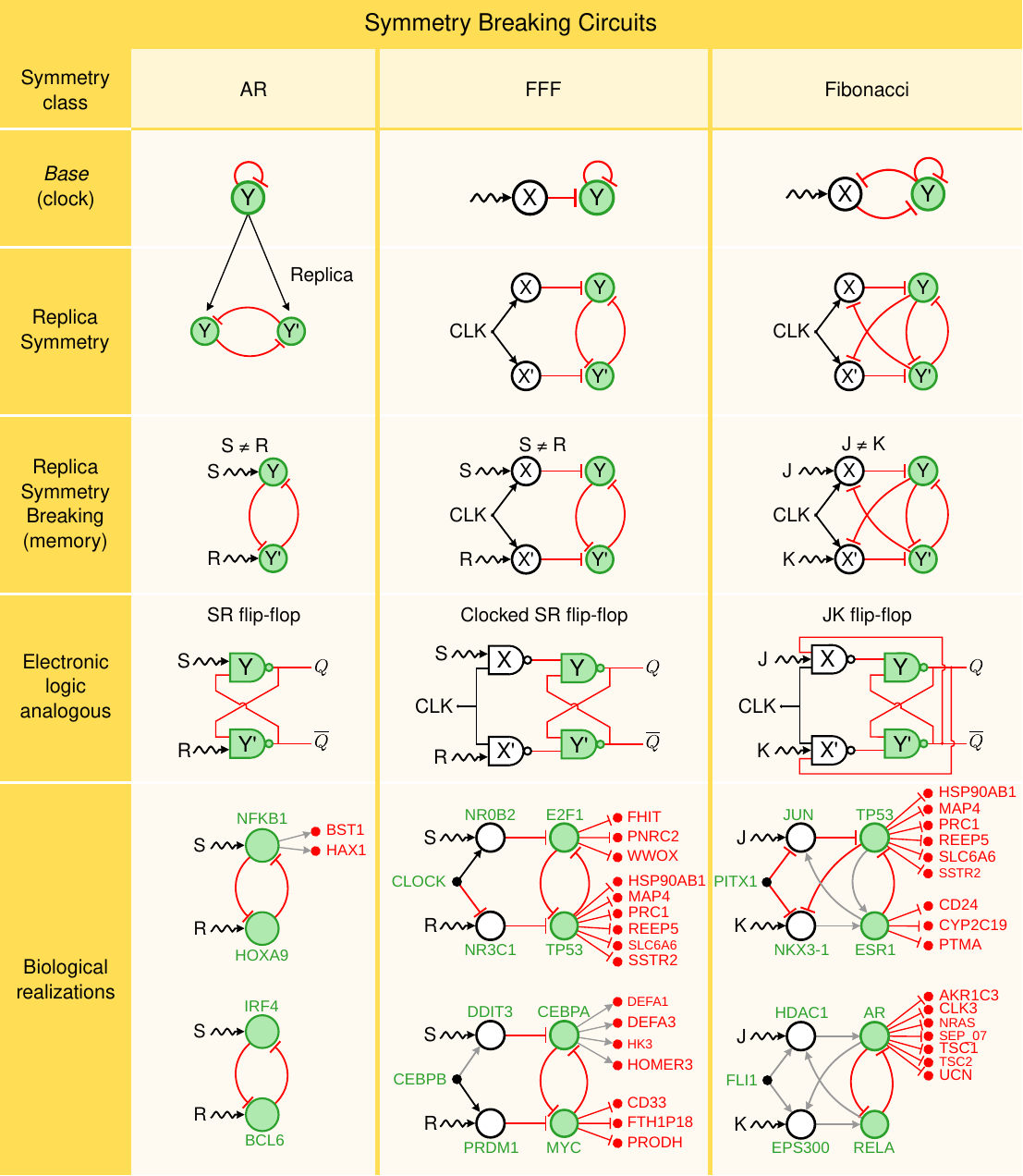}
\caption{\small {\bf Broken symmetry circuits function as memory.} {\bf AR (first column):}  The 
replica symmetry duplication of the AR symmetric circuit results in a network that
is analogous to the SR flip-flop circuit. The symmetry between Y and Y' is
broken by the inputs S and R, such that $\textrm{S} \neq \textrm{R}$, resulting into a pair logical 
outputs $Q$ and $\overline{Q} = $ {\bf not}($Q$).
%
{\bf FFF (second column):} Following the same strategy, we replicate the FFF through a replica 
symmetry duplication. Note that this operation adds a second level of logic gates to the SR 
flip-flop. In order to have consistent logic operations, we add an input clock gene CLK in 
addition to S and R. The resulting circuit is analogous to the Clocked SR flip-flop logic 
circuit.
%
{\bf Fibonacci (third column:} The replica symmetry duplication of the Fibonacci Fiber results in a logic 
circuit which is analogous to the JK flip-flop.
For each symmetry breaking class, we show two examples of circuits from the human
regulatory network. The external regulator genes, S and R (J and K), provide 
inputs which are logically processed by the circuit, accordingly to the type of interaction links
between the genes, activators (black arrows) or repressors (red flat links). 
The outputs of the circuits (green genes) regulate the expression levels of  other genes
(in red) without affecting the circuit functionality. Here, grey arrows correspond to interactions with 
unknown functionality. }
\label{fig4}
\end{figure}

Next, we extend the constructive process described above to include symmetry
breaking.  We do so by mimicking an evolutionary process where
circuits `grow' by a `duplication' event (analogous to gene
duplication in evolution) that conserves the {\it base} of the
original circuit (Fig.~\ref{fig4}, first and second row). Then,
breaking the symmetry creates a new functionality.
We begin this with the simplest case of AR gene Y:
$\left|n=1,\ell=0\right\rangle$. This gene is duplicated by `opening
up' the AR loop into two mutually repressed replica genes, Y and Y' 
(Fig.~\ref{fig4}, second row). This creates a bistable toggle switch
as studied in synthetic genetic circuits~\cite{gardner2000,atkinson2003}
that is topologically equivalent to the core AR loop $|1,0\rangle$ since
both have the same input tree and {\it base} (Fig.~\ref{fig4}, first row).

The symmetry of the replica circuit is then explicitly broken by including
different inputs, S and R, to regulate genes Y and Y', respectively
(Fig.~\ref{fig4}, third row).
S and R represent either genes or effectors that break the symmetry
of the circuit. Symmetry is now broken, and the genetic circuit becomes
analogous to a Set-Reset (SR) flip-flop~\cite{horowitz}, the simplest
canonical circuit to store a bit of binary information in electronic
computer memory (see Fig.~\ref{fig4}, fourth row).

The symmetry is explicitly broken by applying set-reset inputs S~$\ne$~R. 
Specifically, when S~=~1 and R~=~0, the circuit stores a bit of information.
But here is the interesting fact: this state is kept in memory even
when S~=~R~=~1. That is, the circuit remains in the broken symmetry
state even if the inputs are now equal and symmetric. In other words,
the symmetry is now `spontaneously' broken \cite{weinberg}, since a
specific state is selected even without external bias, thus allowing
the circuit to remember its state.

{
Thus, this genetic network is a toggle switch as studied
  in synthetic genetic circuits~\cite{atkinson2003,gardner2000}
  analogous to the SR flip-flop logic circuit shown in
  Fig.~\ref{fig4}, fourth row. When the controlling inputs are $S =0$
  and $R = 1$, the SR flip-flop stores one bit of information
  resulting in $Q = 1$ and $\overline{Q} = 0$ ($\overline{Q} =$ {\bf
    not} $Q$, and the output $Q$ is the logic conversion of {\bf
    not} $Y$). The SR flip-flop retains this logical state even when
  the controlling inputs change. In other words, when $S=1$ and $R=1$,
  the feedback (the repressor interactions between Y and Y') maintains
  the outputs $Q$ and $\overline{Q}$ to its previous state. This state
  changes only when we reset the circuits with $S=1$ and $R=0$. In
  this last case, the SR flip-flop stores $Q=0$ and $\overline{Q} =
  1$, which is also remembered when both inputs are high ($S=1$ and
  $R=1$, S1 File Section~IV~B).  }



This spontaneous symmetry breaking~\cite{weinberg} has analogy
with a ferromagnetic material. When the temperature is
high enough, the ferromagnet does not show any magnetic
property. Moreover, even lowering the temperature, in absence of any
polarizing field, the material does not magnetize. On the contrary, if
an external magnetic field is applied to the ferromagnet, like a magnet put
in contact to a needle for enough time, and then removed, then the needle
becomes magnetic itself, in that it remembers the direction of the previously
applied magnetic field, thus breaking, spontaneously, the rotational symmetry.
Biological realizations of this replica symmetry breaking process are
shown in Fig.~\ref{fig4}, last row. Full list of symmetry broken
circuits in gene regulatory networks across species appears in S1 File
Section~V and S2 File. Table~\ref{asymm_table} shows the Z-scores of these circuits
indicating their significance. The algorithm to identify these broken
symmetry flip-flops in biological networks is developed in S1 File
Section~VII and is available at
\url{https://github.com/makselab/CircuitFinder}.

  \begin{table}[ht]
  \begin{adjustwidth}{-2.5in}{0in} 
  \caption{{\bf Broken symmetry circuits count.}}
   \resizebox{8in}{!}{
      \centering
      \begin{tabular}{| c | c c c | c c c | c c c | c c c |}
        \hline
        Species & Database & Nodes & Edges & & SR flip-flop & & & Clocked SR flip-flop & & & JK flip-flop & \\
        & & & & $N_{\rm real}$ & $N_{\rm rand} \pm SD$ & Z-score & $N_{\rm real}$ & $N_{\rm rand} \pm SD$ & Z-score & $N_{\rm real}$ & $N_{\rm rand} \pm SD$ & Z-score \\
        \hline
        Arabidopsis Thaliana & ATRM & 790 & 1431 & 47 & $1.6 \pm 1.2$ & 36.40 & 3 & $0.2 \pm 0.5$ & 5.80 & 2 & $0 \pm 0$ & $>3$ \\ 
        \hline
        Micobacterium tuberculosis & Research article & 1624 & 3212 & 6 & $1.7 \pm 1.4$ & 3.20 & 0 & N/A & N/A & 0 & N/A & N/A \\ 
        \hline
        Bacillus subtilis & SubtiWiki & 1717 & 2609 & 3 & $2.1 \pm 1.4$ & 0.6 & 0 & N/A & N/A & 0 & N/A & N/A \\ 
        \hline
        Escherichia coli & RegulonDB & 879 & 1835 & 14 & $2.1 \pm 1.4$ & 8.40 & 3 & $0.3 \pm 0.8$ & 3.30 & 0 & N/A & N/A \\ 
        \hline
        Salmonella SL1344 & SalmoNet & 1622 & 2852 & 6 & $1.4 \pm 1.2$ & 3.80 & 0 & N/A & N/A & 0 & N/A & N/A \\ 
        \hline
        Yeast & & & & & 27 & & & 58 & & & 1 & \\
        & YTRP\_regulatory & 3192 & 10947 & 9 & $5 \pm 2.5$ & 1.60 & 3 & $3 \pm 3.6$ & 0 & 0 & N/A & N/A \\
        & YTRP\_binding & 5123 & 38085 & 31 & $21.6 \pm 5.8$ & 1.60 & 192 & $103.3 \pm 45.6$ & 1.90 & 2 & $6.8 \pm 6.1$ & -0.8 \\
        \hline
        Mouse & TRRUST & 2456 & 7057 & 82 & $4 \pm 2.1$ & 37.70 & 216 & $1.9 \pm 2.7$ & 79.50 & 25 & $0.004 \pm 0.06$ & 417 \\ 
        \hline
        Human & & & & & 192 & & & 566 & & & 90 & \\
        & TRRUST & 2718 & 8215 & 89 & $4.3 \pm 2.1$ & 40.50 & 247 & $3.5 \pm 4.8$ & 50.60 & 45 & $0.02 \pm 0.2$ & 225 \\ 
        & TRRUST\_2 & 2862 & 9396 & 103 & $5 \pm 2.3$ & 43 & 319 & $5.9 \pm 7.2$ & 43.20 & 45 & $0.02 \pm 0.3$ & 150 \\ 
        \hline
    \end{tabular}}
    \begin{flushleft} We report the corresponding Z-score statistics as 
    computed in Table~\ref{symm_table}.
\end{flushleft}
    \label{asymm_table}
  \end{adjustwidth}
  \end{table}



Extending this duplication and symmetry breaking process to the FFF
(Fig. \ref{fig4}, second column), one can replicate the X and Y genes
from the FFF {\it base} to create a circuit isomorphic to the Clocked SR
flip-flop, another computer memory building block~\cite{horowitz} (see S1 File
Section~IV B). The symmetry is broken by regulators
or inducers of genes X and X' acting as S (set) and R (reset) of memory,
and it is restored when S~=~R, leaving the system `magnetized'. The hierarchy
continues by replicating the Fibonacci Fiber and consequent breaking of
symmetry when the inputs J and K are different (Fig.~\ref{fig4} third column,
see S1 File Section~IV B). This structure is isomorphic
(i.e., has the same {\it base}) to the JK flip-flop in electronics, which is the
most widely used of all flip-flop designs~\cite{horowitz}.
In its symmetric state, the JK flip-flop is isomorphic to the
symmetric Fibonacci {\it base}. In the symmetry broken phase, it acts
as a memory device which presents two possibilities. A `chiral' symmetry
(where Y feeds X and Y' feeds X') and a `parity' symmetry (left-right reflection,
where Y feeds X' and Y' feeds X). This last one is the one realized in biological
circuits, see Fig.~\ref{fig4}, third column. Examples of JK flip-flops are abundant in gene regulatory networks in human and mouse. They are shown in
Fig.~\ref{fig4}, last row and full list in S1 File Section~V, 
S2 File, and Z-scores in Table~\ref{asymm_table}.





\section*{Discussion}

In summary, fibration symmetries and broken symmetries in gene
regulatory networks reveal the functions of synchronization, clocks
and memory through electronic analogues of transistors, ring
oscillators, current-mirror circuits, and flip-flops. They result in a
hierarchy of building blocks with progressively more complex dynamics
obtained by iterating a procedure of replication and symmetry
breaking.  Beyond the circuits discussed here, the biological
hierarchy can be extended to any number $m$ of loops of length $d$ and
autoregulators in the \textit{fiber} $n$, to form ever more
sophisticated circuits whose complexity is expressed in generalized
Fibonacci sequences $Q_t= n Q_{t-1} + m Q_{t-d}$.

Gene regulatory structures are a mixture of combinational logic
circuits, like FFF, and sequential logic circuits, like FF.  They
provide the network with a structure analogous to a programmable logic
device or chip where the 'register' is a set of flip-flop circuits
tied up together acting as the memory clock of the genetic network
that feeds the combinatorial logic circuits made of simpler
feed-forward circuit of low symmetry. Complex biological circuitry can
then be seen as an emergent process guided by the laws of symmetry
that determine biological functions analogous to electronic
components. The discovery of these building blocks and building rules
of logic computation will allow to: (1) systematically design
synthetic genetic circuits following biological symmetry, and (2)
systematically map the structure and function of all biological
networks, from the symmetries of the connectome ~\cite{morone2019} to
genetic \cite{fibration}, protein and metabolic networks, following a
first principle theoretical approach.

\section*{Supporting information}

%

\paragraph*{S1 File.}
\label{S1_File}
{\bf Supplementary Materials.}  Detailed description of all analytical solutions mentioned in the main text, of the data acquisition and treatment, and detailed description of the proposed algorithm to find fibers.

\paragraph*{S2 File.}
\label{S2_File}
{\bf List of symmetry and broken symmetry circuits.} In S2 File we present a list of circuits found in different species.
%
%
%

\section*{Acknowledgments}
We are grateful to L. Parra and W. Liebermeister for discussions.

\nolinenumbers

%
%
%

\end{document}


\title{Circuits with broken fibration symmetries perform core logic
  computations in biological networks}


\author{Ian Leifer}
\affiliation{Levich Institute and Physics Department, City College
  of New York, New York, New York, USA}
\author{Flaviano Morone}
\affiliation{Levich Institute and Physics Department, City College
  of New York, New York, New York, USA}
\author{Saulo D. S. Reis}
\affiliation{Departamento de F\'isica, Universidade Federal do Cear\'a,
 Fortaleza, Cear\'a, Brazil}
\author{Jos\'e S. Andrade}
\affiliation{Departamento de F\'isica, Universidade Federal do Cear\'a,
Fortaleza, Cear\'a, Brazil}
\author{Mariano Sigman}
\affiliation{Laboratorio de Neurociencia, Universidad Torcuato Di
Tella, Buenos Aires, Argentina}
\author{Hern\'an A. Makse}
\affiliation{Levich Institute and Physics Department, City College
  of New York, New York, New York, USA}

\centerline{ \bf Supplementary Materials:}

\centerline{\bf Circuits with broken fibration symmetries perform core
  logic computations in biological networks}

\centerline{Ian Leifer, Flaviano Morone, Saulo D. S. Reis,
  Jos\'e S. Andrade Jr., Mariano Sigman, and Hern\'an
  A. Makse}
  
\medskip

\tableofcontents

\clearpage

\section{Feed-forward loop: FFL}
\label{sm:ffl}


%
\begin{fmffile}{fflp}
\hspace{-20mm}
\begin{fmfgraph*}(65,24)
\fmftop{i1}
\fmfright{o1,o2}
\fmf{fermion, foreground=blue}{i1,o1}
\fmf{fermion, foreground=blue}{i1,o2}
\fmf{fermion, foreground=blue}{o2,o1}
\fmfv{d.sh=circle, d.f=empty, d.si=.07w, b=(1,,1,,1), l=${\bf X}\ \ \ \ \ \ $}{i1}
\fmfv{d.sh=circle, d.f=empty, d.si=.07w, b=(1,,1,,0), l=${\bf Y}$}{o2}
\fmfv{d.sh=circle, d.f=empty, d.si=.07w, b=(1,,0,,0), l=$\ {\bf Z}$}{o1}
\end{fmfgraph*}
\end{fmffile}
%

{

  In what follows, we analyze in detail the FFL. First, we present the
  analytical solution for the discrete time model of the FFL, where we
  show that the FFL does not synchronize nor oscillates. After that,
  we reach the same conclusion by using a continuous variable
  approach. Finally, we show the discrete time solution with the OR
  gate. Solutions for the FFL have been considered in the
  literature. Here we adapt those results to the particular models
  used in our studies to perform consistent comparisons with the
  solutions of the fiber dynamics obtained through the paper.

\subsection{FFL discrete time model}
\label{ffl-discrete}

Starting from Eq.~(1) in the main text, we define} rescaled variables $\psi_t=y_t/k_y$ and $\zeta_t=z_t/k_y$,
we rewrite  Eq.~(1) as
\begin{equation}
\begin{aligned}
\psi_{t+1} &= \beta\psi_t + \alpha\eta\theta(x_t-k_x),\\
\zeta_{t+1} &= \beta\zeta_t + \alpha\lambda\theta(x_t-k_x)\theta(\psi_t - 1),
\end{aligned}
\label{eq:fflboolrescaled}
\end{equation}
where we set $\beta = (1 - \alpha)$, $\eta = \gamma_x/\alpha k_y$, and
$\lambda = \gamma_x\gamma_y/\alpha k_y$. Equation~(\ref{eq:fflboolrescaled})
defines an iterative map $\psi_{t+1} = f(\psi_t)$ which provides a solution
\begin{equation}
\psi_{t} = f(\psi_{t-1}) = f^2(\psi_{t-2}) = \hdots = f^t(\psi_0).
\end{equation}

A closed form for $\psi_t$ depends on the value of $x$. For the sake
of simplicity, consider $x_t = x$ constant in time. If $x<k_x$, the
solution is simple, and always decays as $\psi_t = \psi_0e^{-t/\tau}$,
where we choose to write $\beta^t = e^{-t/\tau}$ such as $\tau^{-1} =
-\log(1-\alpha)$.
On the other hand, if $x>k_x$, the iterative map is $f(\psi_t) =
\beta\psi_t+\eta$, leading to a solution that converges to $\eta$ as
$\psi_t = \psi_0e^{-t/\tau} +
\eta\left(1-e^{-t/\tau}\right)$. Therefore, the solution to $\psi_t$
is given by:
\begin{equation}
\psi_t = \psi_0e^{-t/\tau}\ + \eta(1-e^{-t/\tau})\theta(x_t-k_x).
\label{eq:fflboolpsisol}
\end{equation}
%
%
%
%

Similarly, the solution for $\zeta_t$ depends on $x$, but it also
depends on $\psi_0$ and $\eta$. For $x<k_x$, it always decays to zero
as $\zeta_t = \zeta_0e^{-t/\tau}$. When $x>k_x$, the variable
$\zeta_t$ follows different behaviors.

For $\psi_0<1$, we also find a
solution that decays as $\zeta_t = \zeta_0e^{-t/\tau}$. However, if
$\eta>1$, this solution ceases to be valid at a given time $t^\ast$
such that $\psi_{t^\ast}>1$, which is given by $t^\ast = \lceil
\tau\log((\eta - \psi_0) / (\eta - 1) )\rceil$. Here, $\lceil x\rceil$
denotes the smallest integer larger than $x$, e.g., $\lceil 1.5\rceil
= 2$. For $t>t^\ast$, as $\psi_t$ saturates to $\eta$, the rescaled
variable $\zeta_t$ converges to $\lambda$ as $\zeta_t =
\zeta_0e^{-(t-t^\ast)/\tau} +
\lambda\left(1-e^{-(t-t^\ast)/\tau}\right)$.

Next, we consider the case $\psi_0>1$. In this case, the solution for $\zeta_t$
is given by $\zeta_t = \zeta_0e^{-t/\tau} + \lambda(1-e^{-t/\tau})$, for $\eta>1$.
In contrast, when $\eta<1$, this solution is valid only up to 
$t^\ast = \lceil\tau\log((\psi_0 - \eta) / (1 - \eta))\rceil$, such that $\psi_t<1$. As
$\psi_t$ saturates to $\eta$, the rescaled variable $\zeta_t$ exponentially decays
as $\zeta_t = \zeta_0e^{-(t - t^\ast)/\tau}$.

To summarize, the possible solutions for $\zeta_t$ depending 
on $\psi_0$ and $\eta$ are:
\begin{equation}
\begin{aligned}
\psi_0>1, \eta >1\ \ \to\ \ \zeta_t &= \zeta_0e^{-t/\tau} +  \lambda(1-e^{-t/\tau}),\\
\psi_0>1, \eta <1\ \ \to\ \ \zeta_t &= \zeta_0e^{-t/\tau} +  \lambda(1-e^{-t/\tau})
\ \ \ {\rm for}\ t\in\Big\{0, 1,\hdots, t_1=\Big\lceil \tau\log \frac{\psi_0-\eta}{1-\eta}\Big\rceil\Big\},\\
\zeta_t &= \zeta_1e^{-(t-t_1)/\tau}\ \ \ {\rm for}\ t>t_1,\\
\psi_0<1, \eta >1\ \ \to\ \ \zeta_t & =\zeta_0e^{-t/\tau}
\ \ \ {\rm for}\ t\in\Big\{0,1, \hdots, t_1=\Big\lceil \tau\log \frac{\eta-\psi_0}{\eta-1}\Big\rceil\Big\},\\
\zeta_t &= \zeta_1e^{-(t-t_1)/\tau}+  \lambda\left(1-e^{-(t-t_1)/\tau}\right)
\ \ \ {\rm for}\ t>t_1,\\
\psi_0<1, \eta <1\ \ \to\ \ \zeta_t &= \zeta_0e^{-t/\tau}. 
\end{aligned}
\end{equation}
Here, $\zeta_1 = \zeta_{t = t_1}$.

From this discussion, we find that the rescaled variables $\psi_t$ and
$\zeta_t$ do not synchronize.  That is, they do not reach the same
value at their fixed points: $\psi_t \neq \zeta_t$ when $t \to
\infty$, unless we use a specific set of parameters.  Moreover,
$\psi_t$ and $\zeta_t$ also do not reach oscillatory states.  The same
conclusion is extended to the original variables $y_t$ and $z_t$.

{
\subsection{FFL ODE}
\label{fff-ode}

In order to show that our results presented in the main text are
consistent with a continuous variable approach, now we focus our
attention on the modelling of the FFL \cite{alon,mangan,alon-delay} by
using ordinary differential equations (ODE). First, we write the ODE
governing the dynamics of expression levels $y(t)$ and $z(t)$:
\begin{equation}
\begin{aligned}
\dot{y}(t) &= -\alpha y(t) + \gamma_x \theta(x(t)-k_x),\\
\dot{z}(t) &= -\alpha z(t) + \gamma_x \theta(x(t)-k_x) \times \gamma_y \theta(y(t)-k_y).
\end{aligned}
\label{eq:fflode}
\end{equation}
For the sake of simplicity, we consider $x(t)=x$ constant in
time.

By using the rescaled functions $\psi(t) = y(t)/k_y$ and $\zeta(t) = k_y$, we rewrite Eq.~(\ref{eq:fflode}) as the following set of ODEs:
\begin{equation}
\begin{aligned}
\dot{\psi}(t) + \alpha\psi(t) &= \alpha\eta\theta(x_t-k_x),\\
\dot{\zeta}(t) + \alpha\zeta(t) &= \alpha\lambda\theta(x_t-k_x)\theta(\psi_t - 1),
\end{aligned}
\label{eq:ffloderescaled}
\end{equation}
where $\eta = \gamma_x/\alpha k_y$ and $\lambda = \gamma_x\gamma_y/\alpha k_y$.

For the case of $x<k_x$, Eqs.~(\ref{eq:ffloderescaled}) become a set
of homogeneous ODEs with solutions that decay exponentially:
\begin{equation}
\begin{aligned}
\psi(t)_{x<k_x} &= \psi_{\rm 0}e^{-\alpha t},\\
\zeta(t)_{x<k_x} &= \zeta_{\rm 0}e^{-\alpha t}.
\end{aligned}
\label{eq:homgenfflodesolution}
\end{equation}
Where, $\psi_{\rm 0}$ and $\zeta_{\rm 0}$ are the initial conditions for the rescaled functions.

When $x>k_x$, we find
\begin{equation}
\psi(t)_{x>k_x} = \psi_{\rm 0}e^{-\alpha t} + \eta \left(1 - e^{-\alpha t}\right).
\end{equation}
Therefore, the solution for $\psi(t)$ converges to $\eta$ as
$t\rightarrow\infty$, similar to the solution of the discrete time
equation presented in the main text.

On the other hand, the solution for $\zeta(t)$ also depends the values
of $\psi_{\rm 0}$ and $\eta$. By carrying on a calculation similar to
the one presented on the main text for the discrete time case, one finds:
\begin{equation}
\begin{aligned}
\psi_0&>1, \eta >1\ \ \to\ \ &\zeta(t)_{x>k_x} &= \zeta_0e^{-\alpha t} + \lambda(1-e^{-\alpha t}),\\
\psi_0&>1, \eta <1\ \ \to\ \ &\zeta(t)_{x>k_x} &= \zeta_0e^{-\alpha t} + \lambda(1-e^{-\alpha t})
\ \ \ {\rm for}\ t\in\left\{0, 1,\hdots, t_1= \frac{1}{\alpha}\log\left(\frac{\psi_0-\eta}{1-\eta}\right)\right\},\\
&&\zeta(t)_{x>k_x} &= \zeta_1e^{-\alpha(t-t_1)}\ \ \ {\rm for}\ t>t_1,\\
\psi_0&<1, \eta >1\ \ \to\ \ &\zeta(t)_{x>k_x} & =\zeta_0e^{-\alpha t}
\ \ \ {\rm for}\ t\in\left\{0,1, \hdots, t_1=\frac{1}{\alpha}\log\left(\frac{\eta-\psi_0}{\eta-1}\right)\right\},\\
&&\zeta(t)_{x>k_x} &= \zeta_1e^{-\alpha(t-t_1)}+  \lambda\left(1-e^{-\alpha(t-t_1)}\right)
\ \ \ {\rm for}\ t>t_1,\\
\psi_0&<1, \eta <1\ \ \to\ \ &\zeta(t)_{x>k_x} &= \zeta_0e^{-\alpha t}. 
\end{aligned}
\end{equation}
Here, $\zeta_1 = \zeta(t)|_{t = t_1}$.

Clearly, from the above solutions, the expression levels from genes Y
and X do not synchronize, since $y(t)\neq z(t)$ as
$t\rightarrow\infty$. In addition, $y(t)$ and $z(t)$ also do not reach
oscillatory states, in accordance with the results of the discrete time 
model.

\subsection{FFL with OR gate}
\label{fff-or}

We present the solution for the FFL~\cite{mangan,alon-delay,alon} with
an OR gate.  We consider the coherent version cFFL where all
regulations are activators. The discrete-time dynamics of expression
levels $y_t$ and $z_t$ of genes Y and Z in the cFFL with a Boolean OR
gate are given by the following pair of difference equations:
\begin{equation}
\begin{aligned}
y_{t+1} &= (1-\alpha) y_t + \gamma_x \theta(x_t-k_x),\\
z_{t+1} &= (1-\alpha) z_t + \gamma_x\theta(x_t-k_x) + \gamma_y\theta(y_t-k_y),
\end{aligned}
\label{eq:ffl-or}
\end{equation}
where $\alpha$, $\gamma_y$, and $\gamma_z$ have the same definition as in the main text. By adopting the same rescaled variables, $\psi_t = y_t/k_y$ and $\zeta_t = z_t/k_y$, we rewrite Eq.~(\ref{eq:ffl-or}) as:
\begin{equation}
\begin{aligned}
\psi_{t+1} &= \beta\psi_t + \alpha\eta\theta(x_t-k_x),\\
\zeta_{t+1} &= \beta\zeta_t + \alpha\lambda_x\theta(x_t-k_x) + \alpha\lambda_x\theta(\psi_t-1),
\end{aligned}
\label{eq:ffl-or-rescaled}
\end{equation}
where we have $\beta = (1-\alpha)$, $\eta = \gamma_x/\alpha k_y$, $\lambda_x = \gamma_x/\alpha k_y$, and $\lambda_y = \gamma_y/\alpha k_y$. Again, without lack of generality, we consider $x_t=x$ constant in time. Clearly, the solution for $\psi_t$ is not affected by the OR gate, therefore, it depends only on the value of $x$ and is given by:
\begin{equation}
\psi_t = \psi_0e^{-t/\tau}\ + \eta(1-e^{-t/\tau})\theta(x_t-k_x).
\label{eq:ffl-or-psisol}
\end{equation}
Again, $\tau = - \log(1-\alpha)$. Thus, if $x<k_x$, $\psi_t$
exponentially decays to zero. On the contrary, if $x>k_x$, $\psi_t$
converges to $\eta$. The solution for $\zeta_t$ displays different
solutions, depending on the combination of $x$ and the initial value
$\psi_{\rm 0}$.

First, consider the case of $x<k_x$. For $\psi_{\rm 0}<1$, we find a
solution that always decays as $\zeta_t = \zeta_{\rm
  0}e^{-t/\tau}$. When $\psi_{\rm 0}>1$, the solution for $\zeta_t$ is
given by $\zeta_t = \zeta_{\rm 0}e^{-t/\tau} + \lambda_y\left(1 -
e^{-t/\tau}\right)$. However, since $x<k_x$ and $\psi_t$ exponentially
decays, this solution is valid only until
$t^\ast~=~\lceil\tau\log\psi_{\rm 0}\rceil$, the instant that
$\psi_{t^\ast}<1$. As $\psi_t$ decays, $\zeta_t$ also decays as
$\zeta_t=\zeta_{t^\ast}e^{-\left(t-t^\ast\right)/\tau}$.

We consider next the case of $x>k_x$. In this case, $\psi_t$ always
converges to $\eta$. Therefore, the solution for $\zeta_t$ depends not
only on $\psi_{\rm 0}$, but also on $\eta$. For $\psi_{\rm 0}<1$ and
$\eta<1$, we find that $\zeta_t$ converges to $\lambda_x$ as $\zeta_t
= \zeta_{\rm 0}e^{-t/\tau} + \lambda_x\left(1 -
e^{-t/\tau}\right)$. However, if $\eta>1$, this solution is not valid
after $t^\ast = \lceil \tau\log\left( \psi_{\rm 0} -
\eta\right)/\left( 1 - \eta \right)\rceil$, when $\psi_t>1$. In this
case, as $\psi_t$ saturates to $\eta$, $\zeta_t$ converges to
$\lambda_x+\lambda_y$ as $\zeta_t = \zeta_{t^\ast}e^{-\left( t-t^\ast
  \right)/\tau} + \left( \lambda_x + \lambda_y \right)\left( 1 -
e^{-\left( t-t^\ast \right)/\tau} \right)$.

For the case of $\psi_{\rm 0}>1$ and $\eta>1$, the rescaled variable
$\zeta_t$ always converges to $\lambda_x + \lambda_y$ as $\zeta_t =
\zeta_{\rm 0}e^{-t/\tau} + \left( \lambda_x + \lambda_y \right)\left(
1 - e^{-t/\tau} \right)$. However, if $\eta<1$, this solution ceases
to be valid at $t^\ast = \lceil \tau\log\left( \left( \eta - \psi_{\rm
  0} \right)/\left( \eta - 1 \right)\right)\rceil$. Therefore, for
this case $\zeta_t$ converges to $\lambda_x$ as
$\zeta_t~=~\zeta_{t^\ast}e^{-\left( t-t^\ast \right)/\tau} +
\lambda_x\left( 1 - e^{-\left( t-t^\ast \right)/\tau}\right)$.

From the discussion above, it is clear that the rescaled variables
$\psi_t$ and $\zeta_t$ neither synchronize nor oscillate. Extending
this results to the original variables $y_t$ and $z_t$, we conclude
that the expression levels of the genes Y and Z also do not
synchronize: $y_t\neq z_t$ when $t\rightarrow\infty$.
}



\section{Satisfied Feed-Forward Fiber: SAT-FFF}
\label{sm:satfff}


\begin{fmffile}{fffp}
\hspace{-20mm}
\begin{fmfgraph*}(65,24)
\fmftop{i1}
\fmfright{o1,o2}
\fmf{fermion, foreground=blue}{i1,o1}
\fmf{fermion, foreground=blue}{i1,o2}
\fmf{fermion, foreground=blue}{o2,o1}
\fmf{fermion, tension=1.8, foreground=blue}{o2,o2}
\fmfv{d.sh=circle, d.f=empty, d.si=.07w, b=(1,,1,,1), l=${\bf X}\ \ \ \ \ \ $}{i1}
\fmfv{d.sh=circle, d.f=empty, d.si=.07w, b=(1,,1,,0), l=${\bf Y}$}{o2}
\fmfv{d.sh=circle, d.f=empty, d.si=.07w, b=(1,,1,,0), l=$\ {\bf Z}$}{o1}
\end{fmfgraph*}
\end{fmffile}


Below, we describe the solutions of the SAT FFF circuit in the
discrete time continuous variable model and in the ODE model.

\subsection{SAT-FFF discrete time model}
\label{sat-discrete}

\begin{figure}[t!]
 \includegraphics[width=\textwidth]{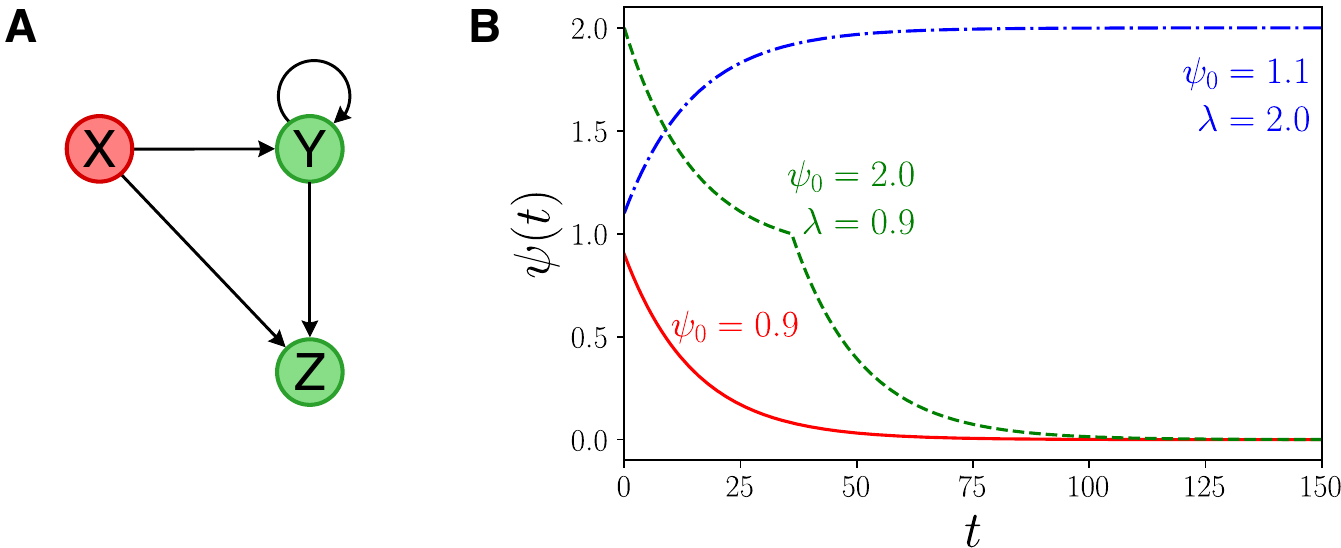}
 \centering
\caption {{\bf SAT-FFF.} {\bf a,} Network representation of the
  SAT-FFF.  All regulations are activators. {\bf b,} Different
behaviors for the analytical solutions of $\psi_t$ depending on
$\psi_0$ and $\lambda$, as discussed in the text.}
\label{sm:fig1}
\end{figure}

The SAT-FFF is a Feed-Forward Fiber with activator autoregulation
where all interactions are satisfied. That is, it does not present the
phenomenon of frustration and the dynamics converges to a fixed
point. This can be simply seen by considering gene X high, which makes
also gene Y high and Z high too.  Finally, the configuration satisfies
the AR loop, so all bonds are satisfied.  The SAT-FFF is constructed
on top of the cFFL by the addition of an autoregulator loop on the Y
gene, as depicted in Fig.~1E (main text) and S1 Fig.~\ref{sm:fig1}A.
The discrete time dynamics of the SAT-FFF with a logic interaction term
 is given by:
\begin{equation}
\begin{aligned}
y_{t+1} &= (1-\alpha)y_t + \gamma_x\theta(x_t-k_x)\times\gamma_y\theta(y_t - k_y),\\
z_{t+1} &= (1-\alpha)z_t + \gamma_x\theta(x_t-k_x)\times\gamma_y\theta(y_t - k_y).
\end{aligned}
\end{equation}
Note that the Heaviside function $\theta(y_t-k_y)$ represents the activator
feedback  on the autoregulation of the Y gene. We consider an AND gate
for the interactions~\cite{alon}. Analogous results can be obtained for 
OR gates or with an first-order ODE model. Writing down the set of equations for the
rescaled variables $\psi_t = y_t/k_y$ and $\zeta_t = z_t/k_y$, we get:
%
\begin{equation}
\begin{aligned}
\psi_{t+1} &= \beta\psi_t + \alpha\lambda\theta(x_t-k_x)\theta(\psi_t-1),\\
\zeta_{t+1} &= \beta\zeta_t + \alpha\lambda\theta(x_t-k_x)\theta(\psi_t-1).
\end{aligned}
\end{equation}
Here, we made use of $\lambda = \gamma_x\gamma_y/\alpha k_y$ and
$\beta = (1-\alpha)$. We note that, since the second term on the right-hand
side of both equations are equal, the dynamical variables $\psi_t$ and $\zeta_t$
must synchronize, as well as $y_t$ and $z_t$. Again, considering $x_t = x$
constant, for $x<k_x$, the solutions for $\psi_t$ and $\zeta_t$ are trivial: both
variables decay exponentially as $\psi_t = \psi_0e^{-t/\tau}$ and
$\zeta_t = \zeta_0e^{-t/\tau}$, where $\tau^{-1} = -\log(1-\alpha)$. This behavior
is shown by the red solid line in S1 Fig.~S\ref{sm:fig1}B with $\psi_0 = 0.9$.

In terms of the iterative map, the dynamics of the SAT-FFF for the
rescaled variable $\psi_t$ with $x>k_x$ is:
\begin{equation}
\psi_{t+1} =  \beta\psi_t +  \alpha\lambda\theta(\psi_t-1) \equiv f(\psi_t),
\label{eq:fffpositiveY}
\end{equation}
so we find
\begin{equation}
f^t(\psi) = f^{t-1}(\beta\psi)\theta(1-\psi) + f^{t-1}(\beta\psi+\lambda)\theta(\psi-1).
\label{eq:fffpositivemap}
\end{equation}
\noindent
This iterative map $\psi_t = f(\psi_t)$ provides different solutions depending
on $\psi_0$.
%
%
%
%
Similar to the case of $x<k_x$, if $\psi_0<1$, the solution decays to zero as
$\psi_{t}= \psi_0e^{-t/\tau}$. However, if $\psi_0>1$, there are two possibilities,
depending on the values of  $\lambda=\gamma_x\gamma_y/\alpha k_y$.

First, if $\lambda>1$, the solution for both rescaled variables converges to
$\lambda$  as $\psi_t=\psi_0e^{-t/\tau} + \lambda(1-e^{-t/\tau})$ and 
$\zeta_t=\zeta_0e^{-t/\tau} + \lambda(1-e^{-t/\tau})$, such that
$\psi_{t\rightarrow\infty}\rightarrow \lambda$ and
$\zeta_{t\rightarrow\infty}\rightarrow \lambda$, as presented by the blue 
dash-dotted line in S1 Fig.~S\ref{sm:fig1}B. For this case, we use
$\psi_0 = 1.1$ and $\lambda = 2$.

For $\lambda<1$, $\psi_t$ approaches
$1$ at a time $t^\ast$  given by 
\begin{equation}
t^\ast = \left\lceil\frac{1}{\log(1 - \alpha)}\log\Bigg(\frac{1-\lambda}{\psi_0 - \lambda}\Bigg)\right\rceil .
\end{equation}
Then, for $t>t^\ast$, the solutions decay to zero as $\psi_t=e^{-(t - t^\ast)/\tau}$
and $\zeta_t = \zeta_{t^\ast}e^{1(t - t^\ast)/\tau}$. This behavior is presented on
S1 Fig.~S\ref{sm:fig1}B by the dashed green line, where we use $\psi_0 = 2$
and $\lambda = 0.9$.
The rescaled variables $\psi_t$ and $\zeta_t$ always synchronize, so
do $y_t$ and $z_t$.  This can be proved by finding the difference
$\epsilon_t = \psi_t - \zeta_t$. For all the cases discussed above, $\epsilon_t$ 
decays exponentially fast as $\epsilon_t = \left(\psi_0 - \zeta_0\right)e^{-t/\tau}$.


Now, we can use the solution with $x_t$ constant to qualitatively
understand the SAT-FFF in general. An example of the SAT-FFF with
non-constant $x_t$ is depicted on Fig.~1F in the main
text. As shown, variable $y_t$ and $z_t$ do synchronize but with no
internal oscillations. We feed an external
oscillatory pattern of $x_t$
as a square wave. For $x_t<k_x$, both $y_t$ an
 $z_t$ decay exponentially. When $x_t>k_x$, they tend to saturate at
 $\gamma_x\gamma_y/\alpha$. The SAT-FFF synchronizes
 at a fixed point.

{
\subsection{SAT-FFF ODE}
\label{sat-ode}

Here, we consider the ODE model of SAT-FFF to confirm results
presented in Section \ref{sat-discrete}.  The dynamics of gene $X$ is
driven by outside sources, so we only consider the dynamics of genes
$Y$ and $Z$ which are described by equations:

\begin{equation}
  \begin{cases}
    \dot{y} = - \alpha y(t) + \gamma_x \theta(x(t)-k_x) \times \gamma_y \theta(y(t)-k_y), \\
    \dot{z} = - \alpha z(t) + \gamma_x \theta(x(t)-k_x) \times \gamma_y \theta(y(t)-k_y). \\
  \end{cases}
  \label{Equations:SAT-FFF-1}
\end{equation}

Taking $\psi(t) = y(t)/k_y$, $\zeta(t) = z(t)/k_y$ and $\delta =
\gamma_x \gamma_y / k_y$ we transform Eq. (\ref{Equations:SAT-FFF-1})
to:

\begin{equation}
  \begin{cases}
    \dot{\psi} = - \alpha \psi(t) + \delta \,\, \theta(x(t)-k_x) \times \theta(\psi(t)-1), \\
    \dot{\zeta} = - \alpha \zeta(t) + \delta \,\, \theta(x(t)-k_x) \times \theta(\psi(t)-1). \\
  \end{cases}
  \label{Equations:SAT-FFF-2}
\end{equation}

Without loss of generality, we consider the case of $x(t)=x$ constant in time. If $x<k_x$, then the solution of Eq. (\ref{Equations:SAT-FFF-2}) is given by:

\begin{equation}
  \begin{cases}
    \psi(t)_{x<k_x} = \psi_{\rm 0}e^{-\alpha t},  \\
    \zeta(t)_{x<k_x} = \zeta_{\rm 0}e^{-\alpha t},
  \end{cases}
  \label{Equations:SAT-FFF-solution_1}
\end{equation}
where $\psi_{\rm 0}$ and $\zeta_{\rm 0}$ are the initial
conditions. Now, let's consider the case $x>k_x$. Equation
(\ref{Equations:SAT-FFF-2}) then transforms into:

\begin{equation}
  \begin{cases}
    \dot{\psi} = - \alpha \psi(t) + \delta \,\, \theta(\psi(t)-1) \\
    \dot{\zeta} = - \alpha \zeta(t) + \delta \,\, \theta(\psi(t)-1) \\
  \end{cases}
  \label{Equations:SAT-FFF-3}
\end{equation}

Due to the existence of isomorphic input trees between both genes,
$\psi(t)$ and $\zeta(t)$ synchronize and therefore $y(t)$ and $z(t)$
synchronize also, so we only consider dynamics of the first equation:

\begin{equation}
    \dot{\psi} = - \alpha \psi(t) + \delta  \,\, \theta(\psi(t)-1) . \\
  \label{Equations:SAT-FFF-4}
\end{equation}

It's easy to see that for $\psi_{\rm 0} < 1$, Eq.
(\ref{Equations:SAT-FFF-solution_1}) will be the solution of Eq.
(\ref{Equations:SAT-FFF-4}). When $\psi_{\rm 0} > 1$ and
$\delta/\alpha > 1$, the solution is given by:

\begin{equation}
    \psi(t) = \delta/\alpha + (\psi_{\rm 0}-\delta/\alpha)e^{-\alpha
      t} .\\
  \label{Equations:SAT-FFF-solution_2}
\end{equation}

In the case $\psi_{\rm 0} > 1$ and $\delta/\alpha < 1$, the dynamics
of $\psi$ is split into two parts. One part before $\psi$ decays to 1
and the other one after $\psi$ crossed 1. The time when $\psi(t)$
crosses 1 is equal to:

\begin{equation}
    t_c = \frac{1}{\alpha} ln(\frac{\psi_{\rm
        0}-\delta/\alpha}{1-\delta/\alpha}),
  \label{Equations:SAT-FFF-critical_time}
\end{equation}
and the dynamics can be written as:

\begin{equation}
  \begin{cases}
    t \in [0, t_c] & \psi(t) = \delta/\alpha + (\psi_{\rm
      0}-\delta/\alpha)e^{-\alpha t}, \\ t \in [t_c, \infty] & \psi(t)
    = \frac{\psi_{\rm 0}-\delta/\alpha}{1-\delta/\alpha} e^{-\alpha
      t}.
  \end{cases}
  \label{Equations:SAT-FFF-solution_3}
\end{equation}

To summarize, the solution is:

\begin{equation}
  \begin{cases}
    \psi_{\rm 0} < 1 & \psi(t) = \psi_{\rm 0}e^{-\alpha t} \\
    \psi_{\rm 0} > 1, \frac{\delta}{\alpha} > 1 & \psi(t) = \delta/\alpha + (\psi_{\rm 0}-\delta/\alpha)e^{-\alpha t} \\
    \psi_{\rm 0} > 1, \frac{\delta}{\alpha} < 1 & 
    \begin{cases}
      t \in [0, t_c] & \psi(t) = \delta/\alpha + (\psi_{\rm 0}-\delta/\alpha)e^{-\alpha t} \\
      t \in [t_c, \infty] & \psi(t) = \frac{\psi_{\rm 0}-\delta/\alpha}{1-\delta/\alpha} e^{-\alpha t}.
    \end{cases}
  \end{cases}
  \label{Equations:SAT-FFF-solutions_summary}
\end{equation}

This solution is analogous to the one obtained for the discrete time
model in Section \ref{sat-discrete}.

}




\section{Unsatisfied Feed-Forward Fiber: UNSAT-FFF}
\label{sm:unsatfff}


\begin{fmffile}{fffn}
\hspace{-20mm}
\begin{fmfgraph*}(65,24)
\fmftop{i1}
\fmfright{o1,o2}
\fmf{fermion, foreground=blue}{i1,o1}
\fmf{fermion, foreground=blue}{i1,o2}
\fmf{fermion, foreground=red}{o2,o1}
\fmf{fermion, tension=1.8, foreground=red}{o2,o2}
\fmfv{d.sh=circle, d.f=empty, d.si=.07w, b=(1,,1,,1), l=${\bf X}\ \ \ \ \ \ $}{i1}
\fmfv{d.sh=circle, d.f=empty, d.si=.07w, b=(1,,1,,0), l=${\bf Y}$}{o2}
\fmfv{d.sh=circle, d.f=empty, d.si=.07w, b=(1,,1,,0), l=$\ {\bf Z}$}{o1}
\end{fmfgraph*}
\end{fmffile}


Now, we turn our attention to the UNSAT-FFF. The solution of this
circuit is developed in the main text using a discrete-time difference
equation with a logistic interaction. Below we elaborate on the
solution of the ODE continuum model and on the conditions on the
period-amplitude relation and the clock functionality.




{

\subsection{UNSAT-FFF ODE}
\label{ode}

The UNSAT-FFF circuit can be reduced to study the base of the circuit
since both genes Y and Z synchronize their behaviour as shown in the main text.
The base of this circuit is a negative autorregulation loop (Fig.~3)
plus an external regulator given by X.  This circuit has been
synthetically implemented by Stricker {\it et al.} in
Ref. \cite{stricker2008} using a promoter that drives the expression
in the absence of LacI (and acts as a negative feedback loop) or in
the presence of IPTG, which acts as an activator. It was shown
experimentally that this circuit leads to oscillatory behaviour in the
expression profiles. This result was corroborated with a dynamical
ODE model in \cite{stricker2008} which here we adapt to study the case
of the UNSAT-FFF with ODE. See also the review paper
\cite{PurcellReview} for further reading.



Following the same approach as above, we consider
gene $x(t)=x$ constant in time and larger than $x>k_x$, and rescale the
expression of genes $y(t)$ and $z(t)$ as $\psi(t) = y(t)/k_y$ and
$\zeta(t) = z(t)/k_z$.  Since genes $y(t)$ and $z(t)$ synchronize
their activities, then only one equation needs to be considered,
$\psi(t)$.

The key to observe oscillations in a first-order ODE is to consider
the delay in the signal propagation in the circuit. Without delay the
dynamics converge to a fixed point; no oscillatory solution exist
in a first-order ODE continuum-time model. The situation is different in the
discrete-time model considered in the text. In this case, a discrete
time plus a logic approximation lead to oscillations.

Negative feedback loop circuits with delays have been widely
investigated in the dynamical systems literature.  Here, we adapt the
negative feedback loop model with delay of Stricker {\it et al.}  (see
Eq. (6) in Supplementary Information in Ref. \cite{stricker2008}). We
consider delays in the negative feedback loop which is the key feature
to explain the experimentally observed robust oscillation in this
circuit \cite{stricker2008}.

Delays in a  biological circuit arise from the combined processes
of intermediate steps like transcription, translation, folding,
multimerization and binding to DNA. This series of biological
processes are lumped into a single arrow between two genes in the
network representation of the circuit. In reality this arrow
represents processes that should be modeled in detail.  These
biological processes can be approximately taken into account by a
delay in the interaction term in the dynamical equations.
The interaction term can be written as $\delta \,\, \theta(1-\psi(t -
\tau))$, where $\tau$ represents the delay caused by the fact that the
process of self-repression is not instant.  Therefore, the dynamics of
$\psi(t)$ are described by a first-order delay-differential equation
(DDE) \cite{stricker2008} of the form:

\begin{equation}
    \dot{\psi} = - \alpha \psi(t) + \delta \,\, \theta(1-\psi(t - \tau)),
  \label{Equations:DDE}
\end{equation}
where $\tau$ represents the delay caused by expression process.


We derive analytical solutions to this equation following a procedure
outlined in \cite{driverDDE} (Chapter V).  We start by noting that
initial conditions used to solve a DDE are not given by the value of
the function at one point, but rather by a set of values of the
function on an interval of length $\tau$. The solution of a DDE can't
be thought of as a sequence of values of $\psi(t)$ as in an ODE, but
rather as a set of functions $\{f_{0}(t), f_{1}(t), f_{2}(t),
\dots,\}$, defined over a set of contiguous time intervals $\{[-\tau,
  0], [0, \tau], [\tau, 2\tau], \dots,\}$.

Let's consider Eq. (\ref{Equations:DDE}) with initial function
$f_0(t)$ for $t \in [-\tau, 0]$. Then for $t \in [0, \tau]$ Eq.
(\ref{Equations:DDE}) looks like:

\begin{equation}
  \dot{\psi} = -\alpha \psi(t) + \delta \,\, \theta(1 - f_{0}(t-\tau)).
  \label{Equations:DDE_general_1}
\end{equation}
Moving the degradation term to the left and multiplying by $e^{\alpha
  t}$ we get:

\begin{equation}
  \dot{\psi}e^{\alpha t} + \alpha \psi(t)e^{\alpha t} = \delta \,\, e^{\alpha t} \theta(1 - f_{0}(t-\tau)).
  \label{Equations:DDE_general_2}
\end{equation}
Re-writing the left part, we obtain:

\begin{equation}
  \frac{d (\psi e^{\alpha t})}{dt} = \delta \,\, e^{\alpha t} \theta(1 - f_{0}(t-\tau)),
  \label{Equations:DDE_general_3}
\end{equation}
and integrating on the interval $\int_{0}^t$, we get:

\begin{equation}
  \psi e^{\alpha t} - \psi(0) = \delta \,\, \int_{0}^t e^{\alpha t'}
  \theta(1 - f_{0}(t'-\tau)) dt \,\,.
  \label{Equations:DDE_general_4}
\end{equation}

Considering that $\psi$ is continuous at 0 ($\psi(0) = f_{0}(0)$) and
$\psi(t)$ for $t \in [0, \tau]$ is given by $f_{1}(t)$:

\begin{equation}
  f_{1}(t) = f_{0}(0) e^{-\alpha t} + \delta \,\, \int_{0}^t e^{\alpha (t'-t)} \theta(1 - f_{0}(t'-\tau)) dt,
  \label{Equations:DDE_general_5}
\end{equation}
then, following the same procedure, we can derive the general formula
for finding the solution $\psi(t)$ on the interval $[k\tau,
  (k+1)\tau]$, assuming that the solution on the previous interval
$[(k-1)\tau, k\tau]$ is given by $f_{k-1}(t)$. We then need to solve
the following iterative equation:

\begin{equation}
  \dot{\psi} = - \alpha \psi(t) + \delta \,\, \theta(1 - f_{k-1}(t-\tau)).
  \label{Equations:DDE_general_6}
\end{equation}

The solution of this equation can be found by applying the integrating
factor method integrating on $\int_{k\tau}^t$. We obtain:

\begin{equation}
  \psi(t) = \psi(k\tau) * e^{\alpha(k\tau-t)} + \delta \,\, \int_{k\tau}^t
  e^{\alpha(t'-t)} \theta(1 - f_{k-1}(t'-\tau)) dt' \, .
  \label{Equations:DDE_general_recursive_equation}
\end{equation}

Using Eq. (\ref{Equations:DDE_general_recursive_equation}) we can
recursively find functions $\{f_{0}(t), f_{1}(t), f_{2}(t), \dots,\}$
on the interval of interest, which provide the solution to Eq.
(\ref{Equations:DDE}). Using Wolfram Mathematica we find functions on
the interval $t \in [-\tau, 30\tau]$ for $f_{0}=2$, $\alpha = 0.2$,
$\delta = 1$ and $\tau=1$ and put them together to find the solution
plotted in S1 Fig.~S\ref{sm:UNSAT-FFF_recursive}A.

We note from Eq.  (\ref{Equations:DDE_general_recursive_equation})
that all functions $f_{k}$ are written as the sum of an exponential
function and a constant.  By looking at Eq.  (\ref{Equations:DDE}), we
see that when the Heaviside function is equal to zero, we get a
solution that decays exponentially to zero. Likewise, when the
Heaviside function is equal to 1, we get a solution that exponentially
grows to $\frac{\delta}{\alpha}$. In other words, the solution will
grow until $\theta(1 - \psi(t-\tau))$ changes to zero (i.e., when
$\psi(t-\tau)>1$) and will decay until $\theta(1 - \psi(t-\tau))$
changes to one (i.e., when $\psi(t-\tau)$ will cross 1 again, but from
different side). Therefore, we get oscillations consisting of two
exponential pieces. One period of the oscillation is shown in
S1 Fig.~S\ref{sm:UNSAT-FFF_recursive}B. The solution on this interval is
given by:

\begin{equation}
  \begin{cases}
    t \in [4.47, 5.69] & \psi(t) = 5 - 10.2 * e^{-0.2t} \\
    t \in [5.69, 9.42] & \psi(t) = 5.4 * e^{-0.2t},
  \end{cases}
  \label{Equations:DDE_oscillation_solution}
\end{equation}
which is the predicted behavior. Additionally, we note that this
circuit functions as a capacitor charging and discharging in an RC
circuit.

\begin{figure}[t!]
 \includegraphics[width=\textwidth]{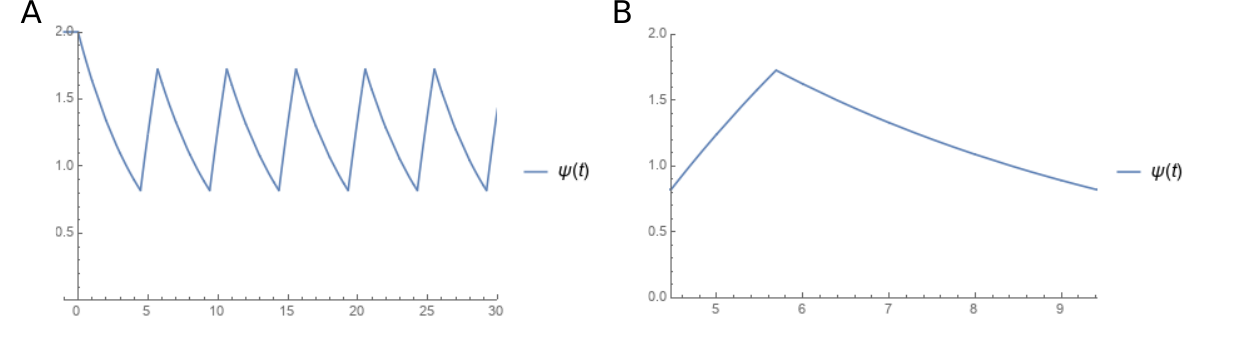}
  \caption {{\bf UNSAT-FFF delay ODE model.} {\bf A,} Solution of
    Eq. (\ref{Equations:DDE}) using recursive Eq.
    (\ref{Equations:DDE_general_recursive_equation}) on $t \in [-\tau,
      30\tau]$ for $f_{0}=2$, $\alpha = 0.2$, $\delta = 1$ and
    $\tau=1$. {\bf B,} One period of the oscillation of solution in
    {\bf (A)} consisting of two exponential pieces.}
  \label{sm:UNSAT-FFF_recursive}
\end{figure}

}

\subsection{Period-amplitude relationship}
As shown in the main text, the solution of the discrete-time Boolean
interaction model for $\lambda>1$ oscillates in time, as well as the
DDE considered in the previous section. Next, we show that this
oscillation has a characteristic amplitude and period. First, to
compute the amplitude of oscillations $A_\psi$ for the rescaled
variable $\psi_t$, we recall that the iterative map $\psi = f(\psi)$
satisfies the recursive equation:
\begin{equation}
f^t(\psi) = f^{t-1}(\beta\psi)\theta(\psi - 1) + f^{t-1}(\beta\psi + \alpha\lambda)\theta(1 - \psi).
\end{equation}
Thus, the amplitude of oscillations $A_\psi$ is given by
\begin{equation}
A_\psi = \lim_{\psi\rightarrow 1^{-}}f(\psi) - \lim_{\psi\rightarrow 1^+}f(\psi) = \alpha\lambda,
\label{eq:amp_psi}
\end{equation}
which implies that
\begin{equation}
A_\psi = \frac{\gamma_x\gamma_y}{k_y}.
\end{equation}

\begin{figure}[t!]
 \includegraphics[width=\textwidth]{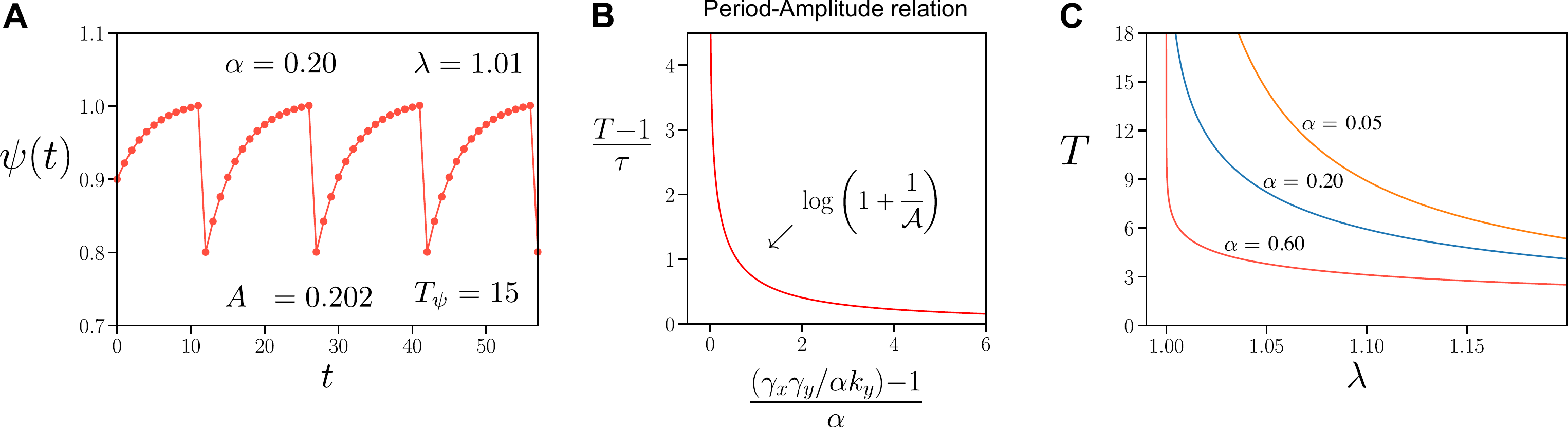}
 \centering
\caption {{\bf Period-amplitude relationship.} {\bf a,} Solutions for $\alpha = 0.2$ and
$\lambda = 1.01$. The values for $A_\psi = 0.202$ and $T = 15$ obtained with the use of 
Eq.~(\ref{eq:amp_psi}) and Eq.~(\ref{eq:period}) perfectly agree with the ones found
by numerical simulations. {\bf b,} Period-amplitude relationship in terms of the original set of 
parameters $\alpha$, $k_y$, $\gamma_x$, and $\gamma_y$. {\bf c,} Period of oscillations
as a function of  $\lambda$ for different values of $\alpha$.}  
\label{sm:fig2}
\end{figure}

To find the period $T$ of the oscillations, we recall from
Eq.~(6) in the main text that the solution for the minimum value
of $\psi$, $\psi_{\rm min}<1$, evolves to its maximum value $\psi_{\rm
  max}$ in $T-1$ iterations as $\psi_{\rm max} =
e^{-(T-1)/\tau}\psi_{\rm min} + \lambda\left(1 -
e^{-(T-1)/\tau}\right)$.  Since $\psi_{\rm min} = (1-\alpha)\psi_{\rm
  max}$, due to the fact that $\psi_{\rm max}>1$, we find

\begin{equation}
T = \left\lceil1 + \tau\log\Bigg(1+\frac{\alpha}{\lambda-1}\Bigg)\right\rceil,
\label{eq:period}
\end{equation}
where we used $\psi_{max} = 1$. For example, using $\alpha = 0.2$ and $\lambda=1.01$,
we find $A_\psi = 2.02$ and $T = 15$, which agrees with the numerical simulation
presented in S1 Fig.~S\ref{sm:fig2}A.

Equation~(\ref{eq:period}) allows to define a rescaled 
amplitude $\mathcal{A} = (\lambda - 1)/\alpha$, and a reduced period $\mathcal{T} = 
(T-1)/\tau$ such that
\begin{equation}
\mathcal{T} = \log\left( 1 + \frac{1}{\mathcal{A}} \right),
\label{eq:period-amp}
\end{equation}
which corresponds to the {\it period-amplitude relationship} of the UNSAT-FFF. A plot
of this relationship is shown in S1 Fig.~S\ref{sm:fig2}B,
where we plot $(T-1)/\tau$
as a function of $\left[(\gamma_x\gamma_y/\alpha k_y)-1\right]/\alpha$.

Coming back to the original variable
$y_t = k_y\psi_t$, we have that the amplitude of oscillations of $y_t$, 
$A = k_yA_\psi$, is given by:

%
%
\begin{equation}
A = \gamma_x\gamma_y,
\label{eq:smamplitude}
\end{equation}
and from Eq.~(\ref{eq:period}), we can write the period of oscillations as
a function of the original set of parameters as
%
\begin{equation}
T = 1 -\frac{1}{\log(1-\alpha)}\log\left( 1 + \frac{\alpha}{(\gamma_x\gamma_y/\alpha k_y) - 1} \right).
\label{eq:smperiod}
\end{equation}
S1 Figure~S\ref{sm:fig2}C shows the period of oscillations $T$ as a function $\lambda$ for
$\alpha = 0.60$, $\alpha = 0.20$, and $\alpha = 0.05$.

\subsection{UNSAT-FFF clock functionality}
\label{sm:function}

The clock functionality of the UNSAT-FFF can be understood by analyzing its
response function, i.e. the relation between oscillations at the input
and at the output of the circuit. The amplitude 
$A_y$, 
and period $T$ of the oscillations are not independent
like in the harmonic oscillator, but are related through a
`period-amplitude' relation expressed by Eq.~(\ref{eq:period-amp})
and S1 Fig.~S\ref{sm:fig2}B. From Eq.~(\ref{eq:smperiod}), for $\alpha$
sufficiently small,
\begin{equation}
T - 1\sim \frac{k_y}{\gamma_x\gamma_y}= \frac{1}{A}\ ,
\label{eq:smpowerperiod}
\end{equation}
which constrains the `clock' ($T$) of the circuit to the power
($A_y$).  As a consequence, $A_y$ and $T$ cannot be controlled
arbitrarily, and this (A-T) constraint helps to stabilize the
UNSAT-FFF response against disturbance in the input X. For example,
for a given available power supply, the system is constrained to
dissipate this power, and when the UNSAT-FFF oscillates, it is
automatically set to operate on an extended time window ($T$ large) at
low amplitude $A$ when a small expression level is required ($A$
small) and vice-versa.  Results for the clock functionality of the
Fibonacci Fiber and $n=2$ Fiber can be carried out in a similar
manner. The idea is that Fibonacci fibers with longer and longer loops
can carry more robust oscillatory patterns than simple autoregulation
negative feedback loops.

\section{Examples of symmetry and broken symmetry circuits}
\label{sm:circuits}

\subsection{Symmetry circuits ({\it fibers}) from \cite{fibration}}
Our findings show that simple sub-graphs ubiquitous on gene regulatory
networks are analogous to symmetric electronic circuits which can work
as clocks, revealing a hierarchy of symmetry circuits.  Here, we
describe these symmetric circuits in more detail following
\cite{fibration}. Supplementary File 1 presents the full list of
circuits found across the regulatory networks of {\it A. thaliana},
{\it M. tuberculosis}, {\it B. subtilis}, {\it E. coli}, {\it
  salmonella}, {\it yeast}, mouse and humans. Below we enumerate the
set of symmetric \textit{fibers} found in Ref. \cite{fibration} in the
genetic networks of these species.

\subsubsection{Repressor regulator link (stub)}
 We start by an isolate repressor link. As depicted in 
Fig.~3A, this repressor regulator alone does not form an input tree, neither it 
forms a {\it base}. Since it works as a transistor, its logic representation is a NOT gate.

\subsubsection{Repressor AR loop: $|1,0\rangle$}
In Fig.~3B, we show the repressor autoregulation loop. When a repressor link is 
found as an AR loop, we have a genetic network with one single loop ($n=1$) and no external 
regulator ($\ell=0$), which we denote symbolically by $|1,0\rangle$. Such simple network has an 
input tree that feeds its own expression levels. Also, it is equivalent to its own {\it base}. By 
the analysis of the corresponding logical circuit, one can observe that it naturally 
oscillates, since it is a NOT gate that feeds itself.

\subsubsection{UNSAT-FFF: $|1,1\rangle$}

The repressor autoregulation loop introduces a symmetry between Y and
Z that allows the expression levels $y_t$ and $z_t$ to synchronize and
oscillate. The increase of external regulators does not affect the
complexity of UNSAT-FFF, since its dynamics remains restricted to the
sole loop on the network. This can be verified in the corresponding
input tree and in its {\it base}. The collapse of Z into Y forms a
{\it base} with $n=1$ autoregulator and $\ell=1$ external
regulator. Because of the repressor feedback, the oscillation and
synchronization of Y and Z are again evident in its logic circuit. In
Fig.~3C, we use the NAND gate for representing gene Y and an
AND gate for gene Z, but other gates (such as an OR logic gate) would
result in similar conclusions.

\subsubsection{Fibonacci Fiber: $\left|1.6180...,0\right\rangle$}
\label{ff-function}

The Fibonacci Fiber shown in Fig.~3D is characterized by the
addition of a second feedback loop. This regulatory network have a
Fibonacci sequence $Q_t = Q_{t-1} + Q_{t-2}$ as an input tree. The
branching structure of the input tree implies that the Fibonacci Fiber
can store memory dynamically by the interaction with its past states,
although it is continually erased in time. As shown in the logic
circuit of Fibonacci Fiber, its structure can oscillate and
synchronize, but is unable to store static information. This situation
changes as soon as we allow symmetries to break, which leads to a
number of genetic circuits, as those depicted in Fig.~4. In
Fig.~3D, we show examples of the Fibonacci Fiber on
regulatory networks of various species. Note the presence of its {\it
  base} in the network representation for {\it B. subtilis} (genes
{\it tnrA} and {\it glnR}), {\it E. coli} (genes {\it uxuR} and {\it
  exuR}), {\it M. tuberculosis} (genes Rv0182c and Rv3286c), {\it
  Salmonella} (genes {\it marA} and {\it soxS}), {\it Yeast} (genes
Tec1 and Ste12), and in two networks from human genetic network (the
pair genes {\it PAX5} and {\it TP53}, for the first example, and gene
{\it FOS} and {\it CREM}, for the second).

\subsubsection{$n=2$ Fiber: $\left|2,0\right\rangle$}

Starting from a Fibonacci Fiber, the addition of a second
autoregulation on gene X results in a symmetric input tree. Besides,
the genetic network of the $n=2$ Fiber collapses into a {\it base}
with a single gene with two autoregulations (Fig.~3E).  From
the corresponding logic circuit, one can conclude that it is possible
to achieve synchronization between genes X, Y and Z. In
Fig.~3E we show examples of the $n=2$ Fiber from the
regulatory networks of {\it B. subtilis} (the first with the pair of
genes {\it lexA} and {\it rocR}, and the second with genes {\it hprT},
{\it tilS} and {\it ftsH}).

\subsection{Symmetry breaking circuits}

Figure~4 shows the procedure to generate broken symmetry
circuits. As we discuss in the main text, this process starts by a
replica symmetry operation where the {\it base} from a given symmetry
circuit is duplicated. The symmetry is broken by the addition of two
input genes as regulators of the circuit.  The resulting broken
symmetry circuits are analogous to flip-flops circuits from digital
electronics. Such circuits are able to store memory statically,
playing a central role in the design of microprocessors. In what
follows, we describe the AR, FFF, and Fibonacci broken symmetry
circuits and their analogous electronic circuits in
detail. Supplementary File 1 presents the full list of circuits found
across the regulatory networks of {\it A. thaliana}, {\it
  M. tuberculosis}, {\it B. subtilis}, {\it E. coli}, {\it
  salmonella}, {\it yeast}, mouse and humans.

\subsubsection{AR symmetry breaking circuit: SR flip-flop} 

Through a replica symmetry duplication, gene Y `opens-up' its AR loop
into two mutually regulated genes Y and Y'. We break the symmetry
relation between Y and Y' by the inclusion of different inputs S and R
as depicted in Fig.~4 (replica symmetry breaking), such that
$\textrm{S} \neq \textrm{R}$. 

The SR flip-flop does not provide synchronized outputs. After the
input signals arrive at the logic gates, each gate provides its output
without waiting for the output of the other. This results in fast
oscillations which, in the particular application to integrated
circuits in digital electronics, are undesired. Then, in digital
electronics, the input $S = 0$ and $R=0$ is said to be forbidden,
since the NAND gates set both $Q=1$ and $\overline{Q}=1$, which
violates the logical state $\overline{Q} =$ {\bf not} $Q$.
In Fig.~4, we use the NAND gates, but the use of a NOR gates
leads to similar conclusions.

Two biological realizations of the AR symmetry breaking class are
shown in Fig.~4, both from human regulatory networks with
genes {\it NFKB1} and {\it HOXA9} (upper), and the regulatory network
of genes {\it IRF4} and {\it BCL6} (bottom). Gene {\it NFKB1} further
regulates two genes, {\it BST1} and {\it HAX1} as its outputs, but
this regulation does not affect the functionality of the flip-flop.

\subsubsection{FFF symmetry breaking circuit: Clocked SR flip-flop}

Following the same strategy, we start with the {\it base} of the UNSAT-FFF and symmetrize it through a 
replica symmetry duplication. Note that the replica symmetry of FFF adds a second level of NAND logic 
gates to the SR flip-flop via gene X. In order to have consistent logic operations, we add an input clock 
gene CLK which symbolizes the activation of gene X, since gene X needs to receive input for its 
activation. The resulting circuit is analogous to the Clocked SR flip-flop after the addition of the input 
genes S and R.  The second level flip-flop inverts the outputs of the previous SR flip-flop logic circuit, 
meaning that when $S=1$ and $R=0$ ($S=0$ and $R=1$), the circuit outputs $Q=1$ and 
$\overline{Q}=0$ ($Q=0$ and $\overline{Q}=1$). The input $S=0$ and $R=0$ results in an 
unchanged state. When the input of gene CLK is low, $CLK = 0$, the output of the second 
level of both NAND gates outputs high signals, independently of the values of S and R, 
assuring that the outputs $Q$ and $\overline{Q}$ remain unchanged. However, when the clock 
input is $CLK=1$, it allows the first level of NAND gates to change the outputs for 
$S\neq R$. The clocked SR flip-flop also has a forbidden state when $S=1$ and $R=1$ in digital 
electronics. In Fig.~4, we show two biological realizations of the FFF broken symmetry breaking 
circuit, the set of genes $\{${\it CLOCK, NR0B2, NR3C1, E2F1} and {\it TP53}$\}$ and the set $\{${\it 
CEBPB,  DDIT3, PRDM1, CEBPA} and {\it MYC}$\}$, both examples are from human regulatory 
genetic  networks. The outputs of the flip-flops, like {\it E2F1} and {\it TP53}, further regulate a set of 
genes each  as  indicated by the red genes in the figure. These regulatory interactions do not affect the 
functionality  of the flip-flops since they are outgoing links.

\subsubsection{Fibonacci symmetry breaking circuit: JK flip-flop}

The replica symmetry duplication of the Fibonacci Fiber results in a logic circuit similar 
to the FFF broken symmetry breaking circuit. However, the regulation links
$Y\rightarrow X'$ and  $Y'\rightarrow X$ yields a different logic circuit, which is 
analogous to the Clocked JK flip-flop, where the input genes are now J and K. The 
additional links solve the unpredictable output for the $J=1$ and $K=1$ case by commuting
the values stored in $Q$ and $\overline{Q}$. Two examples of the Fibonacci broken symmetry 
circuits are shown in Fig.~4 for the sets of genes $\{${\it PITX1, JUN, NKX3-1, TP53} and 
{\it ESR1}$\}$ and  $\{${\it FLI1, HDAC1, EPS300, AR} and {\it RELA}$\}$, both from human genetic 
networks. We also show the set of genes regulated by these flip-flops.

\section{Description of datasets}
\label{sm:datasets} 

Datasets are described in S1 Table \ref{Datasets_table}. We use a set
of datasets of transcriptional regulatory networks found in the
literature. All datasets are freely available from online sources. In
Supplementary File 1 we present a plot of all found circuits. In the
case of symmetric circuits, same colored nodes indicate the genes in
the {\it fiber}.  The external regulators are colored with different
colors. We also present all the symmetry broken circuits across all
species as indicated in the file. The statistics, count and Z-scores
of the circuits are presented in Table~1 and~2 in the main text. The
file with all circuits can be found at \url{https://bit.ly/2YM5x3H}.

\begin{table}[ht!]
    \resizebox{\textwidth}{!}{
	\begin{tabular}{| c | c | p{12cm} |}
      \hline
      Species & Database & Additional information \\\hline
      Arabidopsis Thaliana & ATRM~\cite{ATRM} & We use high-confidence functionally confirmed transcriptional 
      									      regulatory interactions from the ATRM database of the broadly used model plant Arabidopsis. \url{http://atrm.cbi.pku.edu.cn/}
									      \\\hline
      Micobacterium Tuberculosis & Research article~\cite{micobacterium_tuberculosis} &
      									      Supplementary Information of 
									      Ref.~\cite{micobacterium_tuberculosis} \url{https://www.ncbi.nlm.nih.gov/pmc/articles/PMC2600667/bin/msb200863-s2.xls}\\\hline
      Bacillus subtilis & SubtiWiki~\cite{bacillus_subtiwiki} & 
      									     We download the database from SubtiWiki 
									     website and consider all repressor and activation 
									     links as ``Repression" and ``Activation". This database is considered the primary source of information for Bacillus.
									     \url{http://subtiwiki.uni-goettingen.de/}\\\hline
      Escherichia coli & RegulonDB~\cite{regulonDB} &
      			  						     We use the TF - operon interaction network 
									     from~\cite{regulonDB}. RegulonDB combines transcriptional regulator interactions obtained by curating literature and using NLP 
high-quality data and partially confirmed experimentally and computationally predicted data. \url{http://regulondb.ccg.unam.mx/}\\\hline
      Salmonella SL1344 & SalmoNet~\cite{salmonet} & 
      									    We use the regulatory layer of the strain Salmonella 
									    Typhimurium SL1344. SalmoNet consists of manually curated low-throughput and 
high-throughput experiments and predictions based on experimentally verified binding sites 
and TF-gene binding site data from RegulonDB. \url{http://salmonet.org/}\\\hline
      Yeast & YTRP~\cite{YTRP} & 
      									    We use the TF-gene regulatory and TF-gene binding networks.
      									    Results of the TFPEs (Transcription Factor 
Perturbation Experiments) identify the regulatory targets of TFs. This is further
refined by using literature-curated data. \url{http://cosbi3.ee.ncku.edu.tw/YTRP/Home}\\
      \hline
      Mouse & TRRUST~\cite{trrust} & Downloaded from TRRUST website. TRRUST is constructed using sentence-based text mining of more than 
20 million abstracts from research articles, refined by manual curation. \url{https://www.grnpedia.org/trrust/}\\
      \hline
      Human & TRRUST~\cite{trrust} & Downloaded from TRRUST website. \\
      & TRRUST\_2~\cite{trrust} & Downloaded from TRRUST website and curated. \\
      & KEGG~\cite{KEGG} & 
      									    We use KEGG API to download all pathways of
									    Human gene regulatory network. After that, all
									    networks are put together and duplicates are
									    removed. \url{https://www.genome.jp/kegg/pathway.html} \\
      \hline
    \end{tabular}
    }
    \caption{{\bf Description of dataset acquisition.} All data are gathered from publicly available 
    sources~\cite{ATRM,micobacterium_tuberculosis,bacillus_subtiwiki,regulonDB,salmonet,SGD,YTRP,
trrust,KEGG}.}
    \label{Datasets_table}
  \end{table}

  \section{Algorithm to find fibers}
  \label{sm:algorithm}
 
To obtain the set of nodes in the graph that belong to a
\textit{fiber}, we use the algorithm described in detail by Kamei and
Cock~\cite{Kamei} and developed in Ref.~\cite{fibration} to obtain the
`{\it minimal balanced coloring}' of the graph (referred as balanced
coloring for simplicity), where we color the network by assigning a
different color to each \textit{fiber}.

To understand what balanced coloring means in the context of graph
theory, we need to define the concept of input set (which is a part of
the input tree). In a directed graph, the input set of a given node is
the set of nodes with edges pointing into that node. Thus, the input
set is the first layer of the input tree. Next, we define the Input
Set Color Vector (ISCV) of a node as a vector of length equal to the
number of colors in the graph, that is the number of
\textit{fibers}. Each entry of the ISCV of a given node counts how
many nodes of each color are in the input set of this node. The
balanced coloring is achieved by iteration by increasing the number of
colors (length of ISCV) until all nodes of the same color have the
same ISCVs. At this point, each color identifies each \textit{fiber}
in the graph.
  
  Finding graph balanced coloring is equivalent to finding node sets
  with isomorphic input trees. A brief explanation of that is the
  following. If two nodes have the same ISCVs, nodes of their input
  sets have same colors. Thus these nodes are said to belong to the
  same equivalence class~\cite{Kamei}. Inductively, the same can be
  said of the input set of the nodes in the input set. This recurrent
  relation implies that two nodes that have the same input sets will
  have isomorphic input trees and will belong to the same
  \textit{fiber} (for rigorous proof see chapter 4 in \cite{Aldis}).
  
The algorithm to find balanced coloring was described in detail by
Kamei and Cock~\cite{Kamei}. In their algorithm, all nodes of the
graph have the same initial color and, through a series of operations,
they are recolored until balanced coloring is reached. The detailed
description of the algorithm is as follows:
  
  \begin{enumerate}
    \item Initially, all nodes are assigned the same color.
    \item Each node is assigned with the N-dimensional vector (ISCV), where N is current number of 
    colors at the current iteration of the algorithm. Each entry of this vector is the number of nodes of 
    certain color with ingoing link to the respective node. In the first iteration, N = 1 and each entry is the
     in-degree of the node.
    \item If vectors for all nodes of the same color are equal, the balanced coloring is achieved and the 
    algorithm stops.
    \item Otherwise, if coloring is unbalanced, each unique vector is assigned a new color and the graph is recolored accordingly.
    \item Steps 2-4 are repeated until condition in Step 3 is satisfied.
  \end{enumerate}
  
\begin{figure}[h!]
    \includegraphics[width=.8\textwidth]{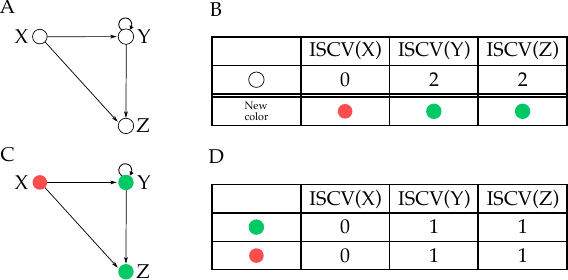} 
    \caption{Illustration of the balanced coloring algorithm to find \textit{fibers}.}
    \label{algorithmDescription}
  \end{figure}

  For example consider the FFF graph in
  S1 Fig.~S\ref{algorithmDescription}A. Initially, we assign the same
  color, white, to all nodes. Then, we assign a 1-dimensional vector
  to each node which counts the in-degree of each node
  (S1 Fig.~S\ref{algorithmDescription}B). Since ISCVs of X and Y (which
  have the same color) are different, ISCV(X)~$=0$ and ISCV(Y)~$=2$,
  where the entry refers to the number of inputs of white color, then
  the condition in Step 3 is not satisfied. There are two unique
  ISCVs, thus only two new colors are necessary. Thus, a 2-dimensional
  vector is assigned to each node: ISCV(X)~$=(0,0)$, ISCV(Y)~$=(1,1)$,
  and ISCV(Z)~$=(1,1)$. Here, the first entry refers to the number of
  inputs of green color and the second entry of red color. Thus, each
  entry of this vector is related to a new color, for example, red and
  green (see S1 Fig.~S\ref{algorithmDescription}D). Then, the network is
  recolored accordingly, as depicted in
  S1 Fig.~S\ref{algorithmDescription}C. At this step, ISCV(Y) and ISCV(Z)
  are the same, and different from the ISCV(X), and both have the same
  color, also different from the color of X, therefore balanced
  coloring is reached and the algorithm stops.  We provide an
  implementation of this algorithm at
  \url{https://github.com/makselab/fiberCodes}.

  \section{Algorithm to find broken symmetry circuits}
  \label{sm:algorithm_broken}
  
\begin{figure}[h!]
    \includegraphics[width=0.7\textwidth]{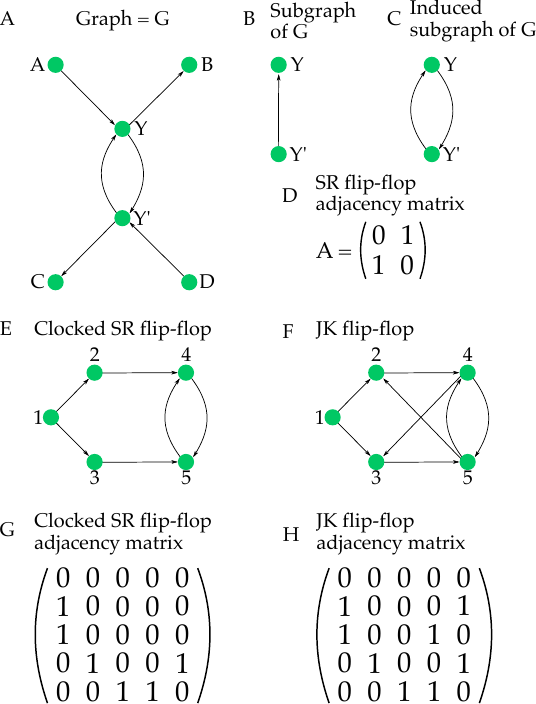} 
    \caption{Definition of the induced subgraph used to obtain the broken symmetry circuits. {\bf A,} Graph G. {\bf B,} Subgraph of G. {\bf C,} Induced Subgraph of G. {\bf D,} Adjacency matrix of SR flip-flop. {\bf E,} Clocked SR flip-flop. {\bf F,} JK flip-flop. {\bf G,} Adjacency matrix of Clocked SR flip-flop. {\bf H,} Adjacency matrix of JK flip-flop.}
    \label{exampleCircuit_SM}
  \end{figure}

To count the number of occurrences of broken symmetry circuits in a
given graph we count the number of appearances of induced subgraphs as
defined in Refs. \cite{subgraph_def, hararyGraphTheory,
  wikiInducedSubgraph}. Subgraphs and induced subgraphs are
graph-theoretical concepts introduced in the social and computer
sciences as applications to graph matching and pattern
recognition~\cite{hararyGraphTheory}.  We allow symmetry breaking to
come from any gene in the circuit.  Let's consider a graph $G = \{V,
E\}$, where $V$ is the set of nodes and $E$ is the set of links. For
instance in S1 Fig.~S\ref{exampleCircuit_SM}A $V = \{A, B, C, D, Y, Y'\}$
and $E = \{A \to Y, Y \to B, Y \to Y', Y' \to Y, Y' \to C, D \to
Y'\}$. A \textit{subgraph} $G' = \{V', E'\}$ of a graph $G = \{V, E\}$
is a graph such that $V' \subset V$ and $E' \subset E$
\cite{subgraph_def}. For example S1 Fig.~S\ref{exampleCircuit_SM}B is a
subgraph of the graph in S1 Fig.~S\ref{exampleCircuit_SM}A with $V' =
\{Y, Y'\} \subset V$ and $E' = \{Y' \to Y\} \subset E$. An
\textit{induced subgraph} $G' = \{V', E'\}$ of a graph $G = \{V, E\}$
is a subgraph with a set of nodes $V' \subset V$ and all links $E'
\subset E$ such that their heads and tails are in $V'$. For example
the subgraph of S1 Fig.~S\ref{exampleCircuit_SM}B is not induced graph of
G since it is missing the link $\{Y \to Y'\}$. That is,
S1 Fig.~S\ref{exampleCircuit_SM}B is not an induced subgraph (but it is
just a subgraph of G), because one of the links $\{Y \to Y'\}$ with
endpoints in $V'$ is missing. However the graph of
S1 Fig.~S\ref{exampleCircuit_SM}C is an induced subgraph of G since $V' =
\{Y, Y'\}$ and $E' = \{Y \to Y', Y' \to Y\}$ are included, but links
$\{A \to Y, Y \to B, Y' \to C, D \to Y'\}$ don't belong to $G'$. The
problem of finding broken symmetry circuits consists on three steps:
(1) identify a {\it base} and create a replica symmetry circuit, (2)
find subgraphs isomorphic to the replica symmetry circuit, and (3)
remove subgraphs that are not induced.
  
  The first step is to find subgraphs isomorphic to the replica
  symmetry circuit. Let's consider an example. The matrix in
  S1 Fig.~S\ref{exampleCircuit_SM}D represents the adjacency matrix $A$
  of the symmetric part of the SR flip-flop (i.e., a toggle
  switch). That is, $A$ is the replica symmetry part (Fig.~4,
  third row) of the broken symmetry circuit. Similarly,
  S1 Fig.~S\ref{exampleCircuit_SM}G,H show the adjacency matrices of
  clocked SR flip-flop and JK flip-flop circuits. The general idea is
  to choose a subgraph and check if it is isomorphic to the circuit
  and continue doing this for all possible subgraphs in the entire
  network. However, this task is computationally expensive.  Even for
  circuits with 5 nodes, the computational time is $~N^5$, where $N$
  is the total number of nodes in $G$, which means that for big enough
  graphs the algorithm can take very long computational
  time. Different approaches to this problem have been widely studied
  and are nicely reviewed in Ref. \cite{graph_matching_review}. Time
  costs can be cut if unprofitable paths are identified and skipped in
  the search space. One of the recent works in the field is the VF2
  algorithm developed by Cordella {\it et al.}~\cite{cordellaVF2}. It
  is designed to deal with large graphs and uses state of the art
  techniques in order to reduce computational time. We use the
  algorithm implemented in a popular R package igraph \cite{igraph} as
  a function $subgraph\_isomorphisms(...)$.  We provide the analysis
  and plotting scripts allowing to reproduce our results at
  \url{https://github.com/makselab/CircuitFinder}.
    
  The second step is to remove all the subgraphs that are not induced
  or, simply speaking, have extra links between the genes in the
  broken symmetry circuit. We follow this procedure: take a node set
  identified above, find the induced subgraph of the complete graph
  with this node set and compare the adjacency matrix of the induced
  subgraph with the adjacency matrix of the circuit. If the matrices
  are different, then the circuit is removed. All remaining circuits
  are the broken symmetry flip-flops that we are looking
  for. Multi-links and self-loops are removed from the network prior
  to consideration.
  
  By applying the steps described above we get the full list of
  induced subgraphs that are isomorphic to the given circuit.

